\documentclass[twocolumn,showpacs,showkeys,preprintnumbers,amsmath,amssymb]{revtex4}
\usepackage{graphicx}% Include figure files
\usepackage{dcolumn}% Align table columns on decimal point
\usepackage{bm}% bold math
\begin{document}
%\preprint{APS/123-QED}
\title{Mass measurements in the vicinity of the rp-process and\\the $\nu$p-process paths with JYFLTRAP and SHIPTRAP}% Force line %breaks with \\
%%%%%%%%%%%%%%%%%%%%%%%%%%%%%%%%%%%%%%%%%%%%%%%%%%%%%%%%%%%%%%%%%%%%%%%%%%%%%%%%%%%%%%%%%%%%%%%%%%
%%%%%%%%%%%%%%%%%%%%%%%%%%%%%%%%%%%%%%%%%%%%%%%%%%%%%%%%%%%%%%%%%%%%%%%%%%%%%%%%%%%%%%%%%%%AUTHORS
%%%%%%%%%%%%%%%%%%%%%%%%%%%%%%%%%%%%%%%%%%%%%%%%%%%%%%%%%%%%%%%%%%%%%%%%%%%%%%%%%%%%%%%%%%%%%%%%%%
\author{C.~Weber$^{1}$}
		\email{christine.weber@phys.jyu.fi}
\author{V.-V.~Elomaa$^{1}$} 
		\thanks{This publication comprises part of the Ph.D. thesis of V.-V.~Elomaa.}
\author{R.~Ferrer$^{2}$}
		\thanks{This publication comprises part of the Ph.D. thesis of R. Ferrer.}
		\altaffiliation{Present address: NSCL, Michigan State University, East Lansing, MI 48824-1321, USA.}
\author{C.~Fr\"ohlich$^{3}$}
\author{D.~Ackermann$^{4}$}
\author{J.~\"Ayst\"o$^{1}$}
\author{G.~Audi$^{5}$}
\author{L.~Batist$^{6}$}
\author{K.~Blaum$^{2,4}$}
    \altaffiliation{Present address: Max-Planck-Institut f\"ur Kernphysik, D-69117 Heidelberg, Germany.}
\author{M.~Block$^{4}$}
\author{A.~Chaudhuri$^{7}$}
    \altaffiliation{Present address: Department of Physics and Astronomy, University of Manitoba, Winnipeg, MB, R3T2N2, Canada.}
\author{M.~Dworschak$^{4}$}
\author{S.~Eliseev$^{4,6}$}
    \altaffiliation{Present address: Max-Planck-Institut f\"ur Kernphysik, D-69117 Heidelberg, Germany.}
\author{T.~Eronen$^{1}$}
\author{U.~Hager$^{1}$}
	  \altaffiliation{Present address: TRIUMF, 4004 Wesbrook Mall, Vancouver BC, V6T 2A3, Canada.}
\author{J.~Hakala$^{1}$}
\author{F.~Herfurth$^{4}$}
\author{F.P.~He{\ss}berger$^{4}$}
\author{S.~Hofmann$^{4}$}
\author{A.~Jokinen$^{1}$}
\author{A.~Kankainen$^{1}$}
\author{H.-J.~Kluge$^{4,8}$}
\author{K.~Langanke$^{4}$}
\author{A.~Mart\'{\i}n$^{4}$}
\author{G.~Mart\'{\i}nez-Pinedo$^{4}$}
\author{M.~Mazzocco$^{4}$}
			\altaffiliation{Present address: Dipartimento di Fisica and INFN - Sezione di Padova, via Marzolo 8, I-35131 Padova, Italy.}
\author{I.D.~Moore$^{1}$}
\author{J.B.~Neumayr$^{9}$}
\author{Yu.N.~Novikov$^{4,6}$}
\author{H.~Penttil\"a$^{1}$}
\author{W.R.~Pla{\ss}$^{4,10}$}
\author{A.V.~Popov$^{6}$}
\author{S.~Rahaman$^{1}$}
\author{T.~Rauscher$^{11}$}
\author{C.~Rauth$^{4}$}
\author{J.~Rissanen$^{1}$}
\author{D.~Rodr\'{\i}guez$^{12}$}
\author{A.~Saastamoinen$^{1}$}
\author{C.~Scheidenberger$^{4,10}$}
\author{L.~Schweikhard$^{7}$}
\author{D.M.~Seliverstov$^{6}$}
\author{T.~Sonoda$^{1}$}
		\altaffiliation{Present address: Atomic Physics Laboratory, RIKEN, 2-1 Hirosawa, Wako, Saitama 351-0198, Japan.}
\author{F.-K.~Thielemann$^{11}$}
\author{P.G.~Thirolf$^{9}$}
\author{G.K.~Vorobjev$^{4,6}$}
%%%%%%%%%%%%%%%%%%%%%%%%%%%%%%%%%%%%%%%%%%%%%%%%%%%%%%%%%%%%%%%%%%%%%%%%%%%%%%%%%%%%%%%%%%%%%%%%%%
%%%%%%%%%%%%%%%%%%%%%%%%%%%%%%%%%%%%%%%%%%%%%%%%%%%%%%%%%%%%%%%%%%%%%%%%%%%%%%%%%%%%%%%%%%%%%%%%%%
\affiliation
{$^{1}$Department of Physics, University of Jyv\"askyl\"a, FI-40014 Jyv\"askyl\"a, Finland}
\affiliation
{$^{2}$Institut f\"ur Physik, Johannes Gutenberg-Universit\"at, D-55099 Mainz, Germany}
\affiliation
{$^{3}$The Enrico Fermi Institute, Department of Astronomy and Astrophysics, The University of Chicago, Chicago, IL 60637, USA}
\affiliation
{$^{4}$GSI-Darmstadt mbH, D-64291 Darmstadt, Germany}
\affiliation
{$^{5}$CSNSM-IN2P3/CNRS, Universit\'{e} de Paris-Sud, F-91405 Orsay, France}
\affiliation
{$^{6}$Petersburg Nuclear Physics Institute, 188300 Gatchina, St. Petersburg, Russia}
\affiliation
{$^{7}$Institut f\"ur Physik, Ernst-Moritz-Arndt-Universit\"at, D-17487 Greifswald, Germany}
\affiliation
{$^{8}$Ruprecht-Karls-Universit\"at Heidelberg, D-69120 Heidelberg, Germany}
\affiliation
{$^{9}$Fakult\"at f\"ur Physik, Ludwig-Maximilians-Universit\"at M\"unchen, D-85748 Garching, Germany}
\affiliation
{$^{10}$II. Physikalisches Institut, Justus-Liebig-Universit\"at, D-35392 Gie{\ss}en, Germany}
\affiliation
{$^{11}$Departement f\"ur Physik, Universit\"at Basel, CH-4056 Basel, Switzerland}
\affiliation
{$^{12}$Universidad de Huelva, Avda. de las Fuerzas Armadas s/n, E-21071 Huelva, Spain}
%%%%%%%%%%%%%%%%%%%%%%%%%%%%%%%%%%%%%%%%%%%%%%%%%%%%%%%%%%%%%%%%%%%%%%%%%%%%%%%%%%%%%%%%%%%%%%%%%
%%%%%%%%%%%%%%%%%%%%%%%%%%%%%%%%%%%%%%%%%%%%%%%%%%%%%%%%%%%%%%%%%%%%%%%%%%%%%%%%%%%%%%Abstract
%%%%%%%%%%%%%%%%%%%%%%%%%%%%%%%%%%%%%%%%%%%%%%%%%%%%%%%%%%%%%%%%%%%%%%%%%%%%%%%%%%%%%%%%%%%%%%%%%%
\date{\today}%  It is always \today, today,
\begin{abstract}The masses of very neutron-deficient nuclides close to the astrophysical rp- and 
$\nu$p-process paths have been determined with the Penning trap facilities JYFLTRAP at 
JYFL/Jyv\"askyl\"a and SHIPTRAP at GSI/Darmstadt. Isotopes from yttrium ($Z = 39$) to palladium 
($Z = 46$) have been produced in heavy-ion fusion-evaporation reactions. In total 21 nuclides were 
studied and almost half of the mass values were experimentally determined for the first time: 
$^{88}\mbox{Tc}$, $^{90-92}\mbox{Ru}$, $^{92-94}\mbox{Rh}$, and $^{94,95}\mbox{Pd}$. For the $^{95}\mbox{Pd}^{m}$, 
($21/2^+$) high-spin state, a first direct mass determination was performed. Relative mass 
uncertainties of typically $\delta m / m = 5 \times 10^{-8}$ were obtained. The impact of the new mass 
values has been studied in $\nu$p-process nucleosynthesis calculations. The resulting reaction flow 
and the final abundances are compared to those obtained with the data of the Atomic Mass 
Evaluation 2003.  
\end{abstract}
\pacs{07.75.+h Mass spectrometers, 21.10.Dr Binding energies and
masses, 26.30.-k Nucleosynthesis in novae, supernovae, and other explosive environments, 
26.50.+x Nuclear physics aspects of novae, supernovae, and other explosive environments}
\keywords{atomic mass, nucleosynthesis, $\nu$p-process, reaction rates, $^{84}\mbox{Y}$,
$^{87}\mbox{Zr}$, $^{88}\mbox{Mo}$, $^{89}\mbox{Mo}$,
$^{88}\mbox{Tc}$, $^{89}\mbox{Tc}$, $^{90}\mbox{Tc}$,
$^{91}\mbox{Tc}$, $^{92}\mbox{Tc}$, $^{90}\mbox{Ru}$,
$^{91}\mbox{Ru}$, $^{92}\mbox{Ru}$, $^{93}\mbox{Ru}$,
$^{94}\mbox{Ru}$, $^{92}\mbox{Rh}$, $^{93}\mbox{Rh}$,
$^{94}\mbox{Rh}$, $^{95}\mbox{Rh}$, $^{94}\mbox{Pd}$,
$^{95}\mbox{Pd}$, $^{95}\mbox{Pd}^{m}$, $^{96}\mbox{Pd}$
determined with the Penning trap mass spectrometers JYFLTRAP and
SHIPTRAP} 
\maketitle
%%%%%%%%%%%%%%%%%%%%%%%%%%%%%%%%%%%%%%%%%%%%%%%%%%%%%%%%%%%%%%%%%%%%%%%%%%%%%%%%%%%%%%%%%%%%%%%%%%
%%%%%%%%%%%%%%%%%%%%%%%%%%%%%%%%%%%%%%%%%%%%%%%%%%%%%%%%%%%%%%%%%%%%%%%%%%%%%%%%%%%%%%INTRODUCTION
%%%%%%%%%%%%%%%%%%%%%%%%%%%%%%%%%%%%%%%%%%%%%%%%%%%%%%%%%%%%%%%%%%%%%%%%%%%%%%%%%%%%%%%%%%%%%%%%%%
\section{\label{intro}Introduction}
Direct mass measurements of exotic nuclei have advanced considerably due to the introduction of Penning-trap-based 
approaches at on-line facilities \cite{Stol1990,Blau2006,IJMS2006}. At present, relative uncertainties on 
the order of $5 \times 10^{-8}$ or better are routinely achieved. These results provide important contributions towards studies 
of nuclear structure evolution far from stability. In particular, the high accuracy can play a crucial role in 
investigations of nuclei near the $N = Z$ line in the form of precision tests of the weak interaction \cite{Blau2003a,Eron2008,Town2008}, 
charge-symmetry effects in nuclear structure \cite{Warn2006}, and in nuclear astrophysics \cite{Scha1998,Brow2002}.\\
In this paper we report on accurate mass measurements of neutron-deficient yttrium to palladium isotopes to ascertain their 
impact on the nucleosynthesis of neutron-deficient isotopes below tin. In this region, the rapid proton-capture process 
(rp-process) runs along the $N = Z$ line as a sequence of (p,$\gamma$) reactions followed by $\beta^{+}$ decays. This process 
is associated with accreeting neutron stars where explosive hydrogen burning gives rise to the observed X-ray bursts. For 
very exotic nuclides close to the proton drip line, direct reaction-rate measurements are hampered by low production cross 
sections. Consequently, the necessary reaction rates have to be evaluated by theoretical models where the nuclear masses are 
one of the key input quantities. For low enough temperatures the rates are determined by individual resonances and depend 
exponentially on the resonance energy \cite{Scha2006}. For higher temperatures many resonances are located in the Gamow 
window and therefore statistical models are applicable.\\
While the rp-process is necessary to explain the observed X-ray bursts, its contribution to the solar abundances is doubtful 
as the produced nuclei are not ejected from the neutron star. In the same mass region the $\nu$p-process 
\cite{Froh2006,Prue2006,Wana2006} occurs as sequence of (p,$\gamma$) and (n,p) or $\beta^+$ reactions producing neutron-deficient 
nuclei with $A > 64$. This process occurs in proton-rich supernova ejecta under the influence of strong neutrino and 
antineutrino fluxes which produce neutrons via antineutrino absorption on free protons. The (n,p) reactions allow to overcome 
the long $\beta$-decay half-lives. After the temperature drops, proton-capture reactions freeze out and matter decays back to the 
line of $\beta$ stability. In this way, neutron-deficient nuclei are produced and ejected in the supernova explosion.\\
%%%%%%%%%%%%%%%%%%%%%%%%%%%%%%%%%%%%%%%%%%%%%%%%%%%%%%%%%%%%%%%%%%%%%%%%%%%%%%%%%%%%%%%%%%%%%%%%%%
%%%%%%%%%%%%%%%%%%%%%%%%%%%%%%%%%%%%%%%%%%%%%%%%%%%%%%%%%%%%%%%%%%%%%%%%%%%%%%%%%%%%%%EXPERIMENTAL
%%%%%%%%%%%%%%%%%%%%%%%%%%%%%%%%%%%%%%%%%%%%%%%%%%%%%%%%%%%%%%%%%%%%%%%%%%%%%%%%%%%%%%%%%%%%%%%%%%
\section{\label{exp}Experimental setup and procedure}
\begin{figure}
\includegraphics[angle=90,width=0.48\textwidth]{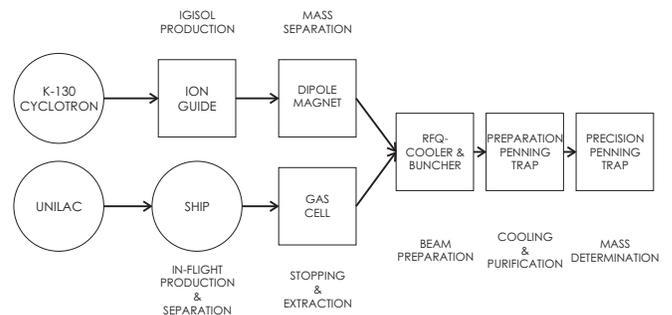}
\caption{\label{EXP_LAYOUTS}Functional layout of the Penning trap
mass spectrometers JYFLTRAP/JYFL (top) and SHIPTRAP/GSI (bottom). For a
detailed description see text.}
\end{figure}

\begin{figure*}
\includegraphics[width=0.9\textwidth]{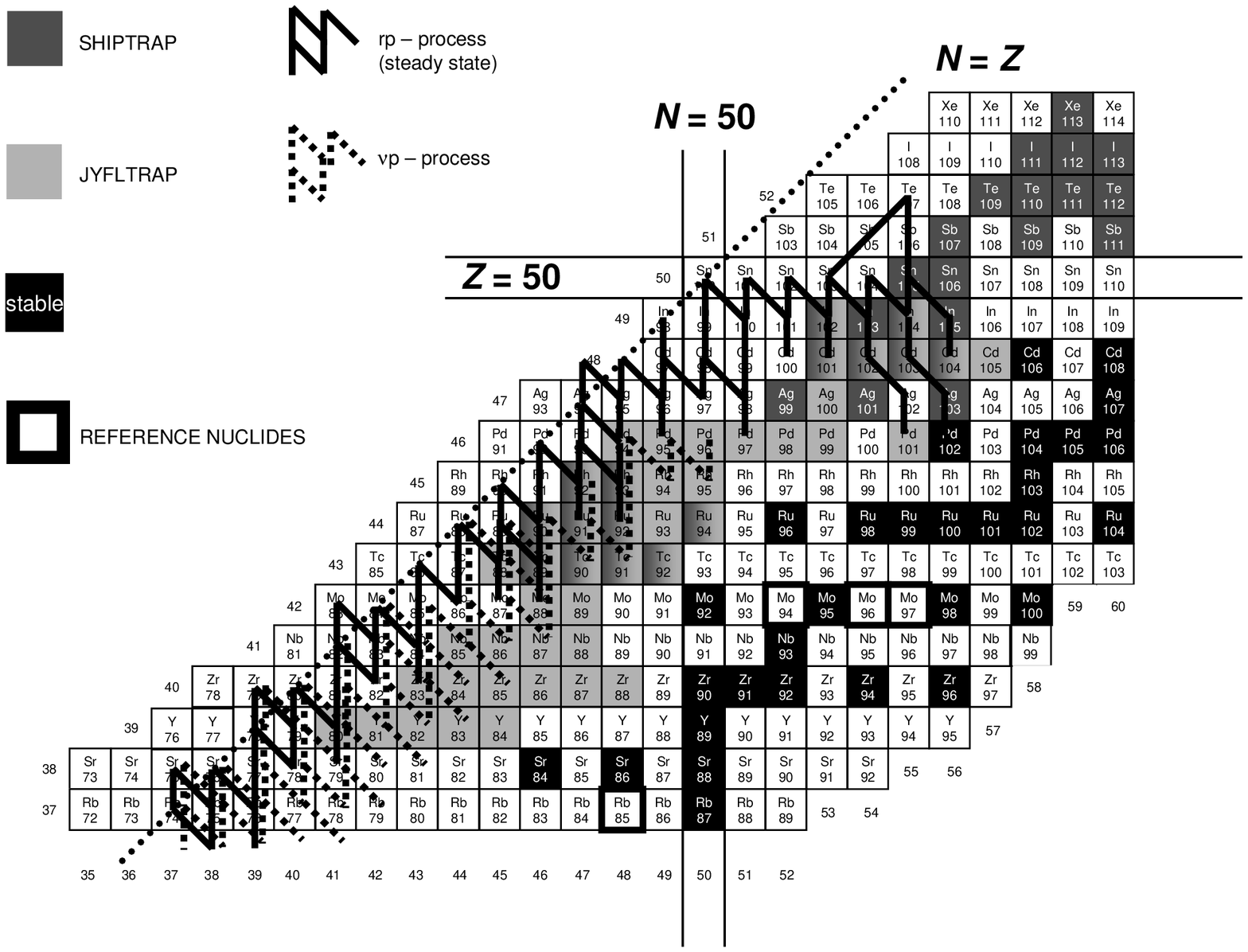}
\caption{\label{chart_JYFL_SHIP}Nuclides studied at JYFLTRAP (light
grey) and SHIPTRAP (dark grey). The present paper reports on
masses of $^{84}\mbox{Y}$, $^{87}\mbox{Zr}$ and isotopes of
molybdenum, technetium, ruthenium, rhodium, and
$^{94-96}\mbox{Pd}$. All other nuclides measured by either trap facility are published 
in Refs. \cite{Kank2006,Mart2007,Elom2008}. The $N = Z$ line was reached in 
measurements with the mass spectrometer ISOLTRAP for rubidium \cite{Kell2007,Otto1994} 
and strontium isotopes \cite{Sikl2005,Otto1994}.}
\end{figure*}

The data presented in this paper were obtained within joint experiments at two facilities: SHIPTRAP at GSI and 
JYFLTRAP in Jyv\"askyl\"a. Although the stopping and separation mechanisms for short-lived, exotic species 
are different at each facility, the basic functional units of the Penning trap setups are identical. 
Figure \ref{EXP_LAYOUTS} shows the schematic layout for both Penning trap mass spectrometers, indicating their 
differences and similarities.\\
The IGISOL facility provides nuclides after bombarding a thin target with a beam from the Jyv\"askyl\"a 
K-130 cyclotron. The reaction products recoil into helium gas at pressures of typically 
$p \approx 200~\mbox{mbar}$ and thermalize therein, ending with a high fraction as singly-charged ions 
\cite{Ayst2001}. After extraction by the helium flow and electric field guidance, the ions pass through a dipole 
magnet at an energy of $30 - 40~\mbox{keV}$, where they are mass-separated with a resolving power of up to 
$R = 500$ and finally delivered to the JYFLTRAP system. One of the main advantages of this chemically 
non-selective method is the availability of refractory elements.\\
The SHIPTRAP facility \cite{Dill2000,Bloc2005} is located behind the velocity filter SHIP \cite{Munz1979,Hofm2000} 
at GSI/Darmstadt. Here, short-lived nuclides are produced by fusion-evaporation reactions in a thin target 
and are separated from the primary beam in a double Wien filter. The major physics goals of this experimental 
setup are the studies of transuranium elements, which cannot be accessed at other trap facilities. The reaction 
products from SHIP with energies of a few $100~\mbox{keV/u}$ are stopped in a buffer-gas filled stopping cell 
\cite{Neum2006}. The ions are extracted from the gas cell by a combination of DC- and RF-electric fields 
through a nozzle. Subsequently, they are pre-cooled in an extraction radiofrequency quadrupole (RFQ)
operated as an ion guide.\\
In both experiments, exotic ion beams are produced, decelerated, cooled and purified prior to their mass 
determination in the Penning trap. The extracted ion beam enters a gas-filled radiofrequency quadrupole (RFQ) 
cooler and buncher \cite{Niem2001,Herf2001} where the continuous beam is cooled and bunched for an efficient 
injection into the respective Penning trap system. These are nearly identical and have been developed in a 
common effort \cite{Sikl2003,Kolh2003,Kolh2004}. Both traps are open-endcap, cylindrical Penning traps 
\cite{Gabr1989}, with a 7-electrode electrode configuration \cite{Raim1997} for the preparation trap. The 
second trap, for the mass determination, is of identical geometry as the first one at JYFLTRAP, whereas at 
SHIPTRAP a 5-electrode trap with a reduced storage region is employed. They are placed in the warm bore of the 
same type of superconducting solenoid with two homogeneous centers at a field strength of $B = 7~\mbox{T}$. The
relative inhomogeneities of the magnetic fields $\Delta B/ B$ within $1~\mbox{cm}^3$ are close to $1~\mbox{ppm}$ 
and $0.1~\mbox{ppm}$ at the positions of the first and the second trap, respectively.\\
In the first Penning trap a mass-selective buffer-gas cooling scheme \cite{Sava1991} is employed in order to 
separate individual nuclides. The resolving power, 
$R = \nu_{\mbox{\scriptsize c}}/\Delta \nu_{\mbox{\scriptsize c,FWHM}} = m/\Delta m$, that can be
obtained in this process reaches up to $R \approx 1 \times 10^5$.
\begin{figure}
\includegraphics[width=0.48\textwidth]{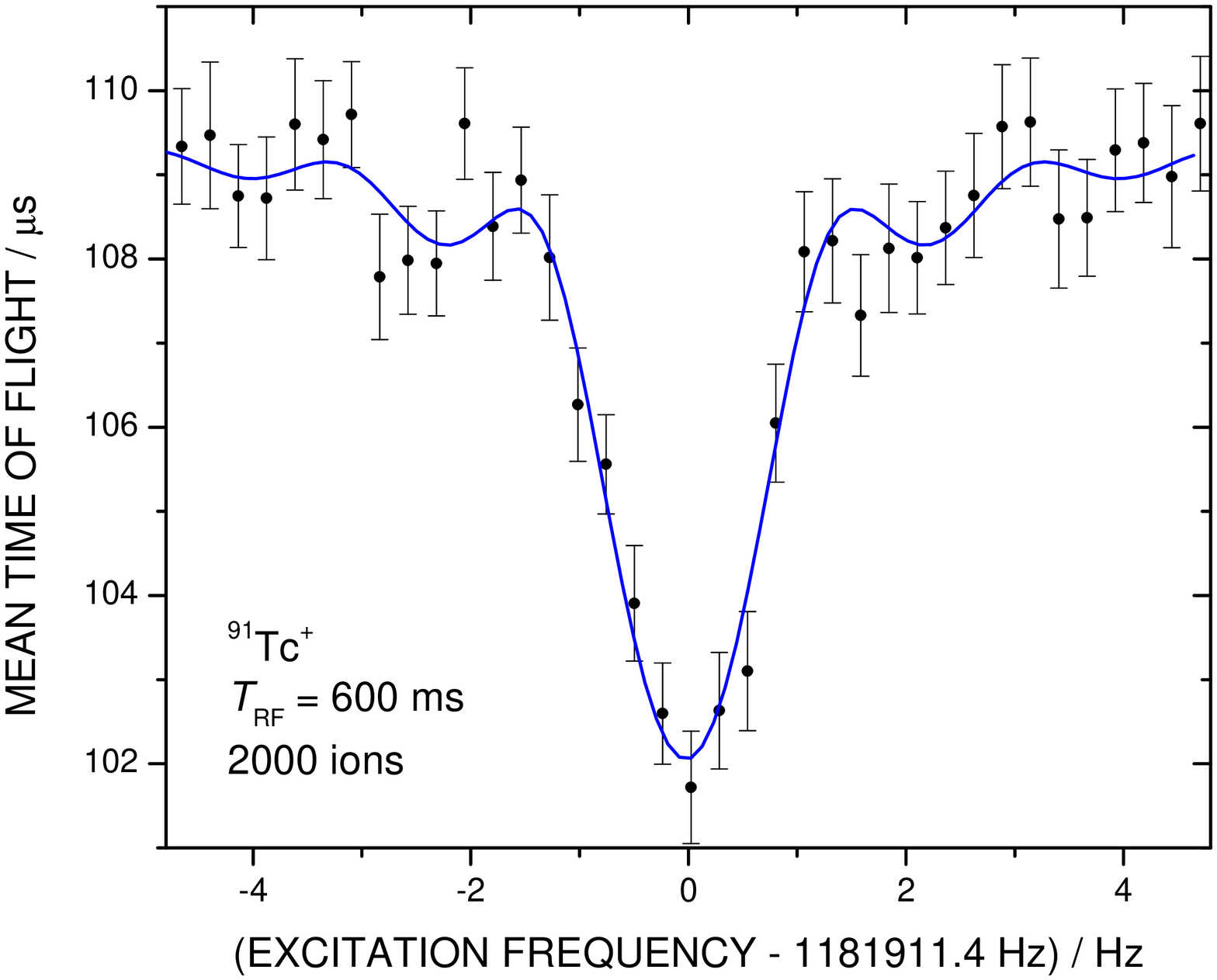}
%\end{figure}
\\[20pt]
%\begin{figure}
\includegraphics[width=0.48\textwidth]{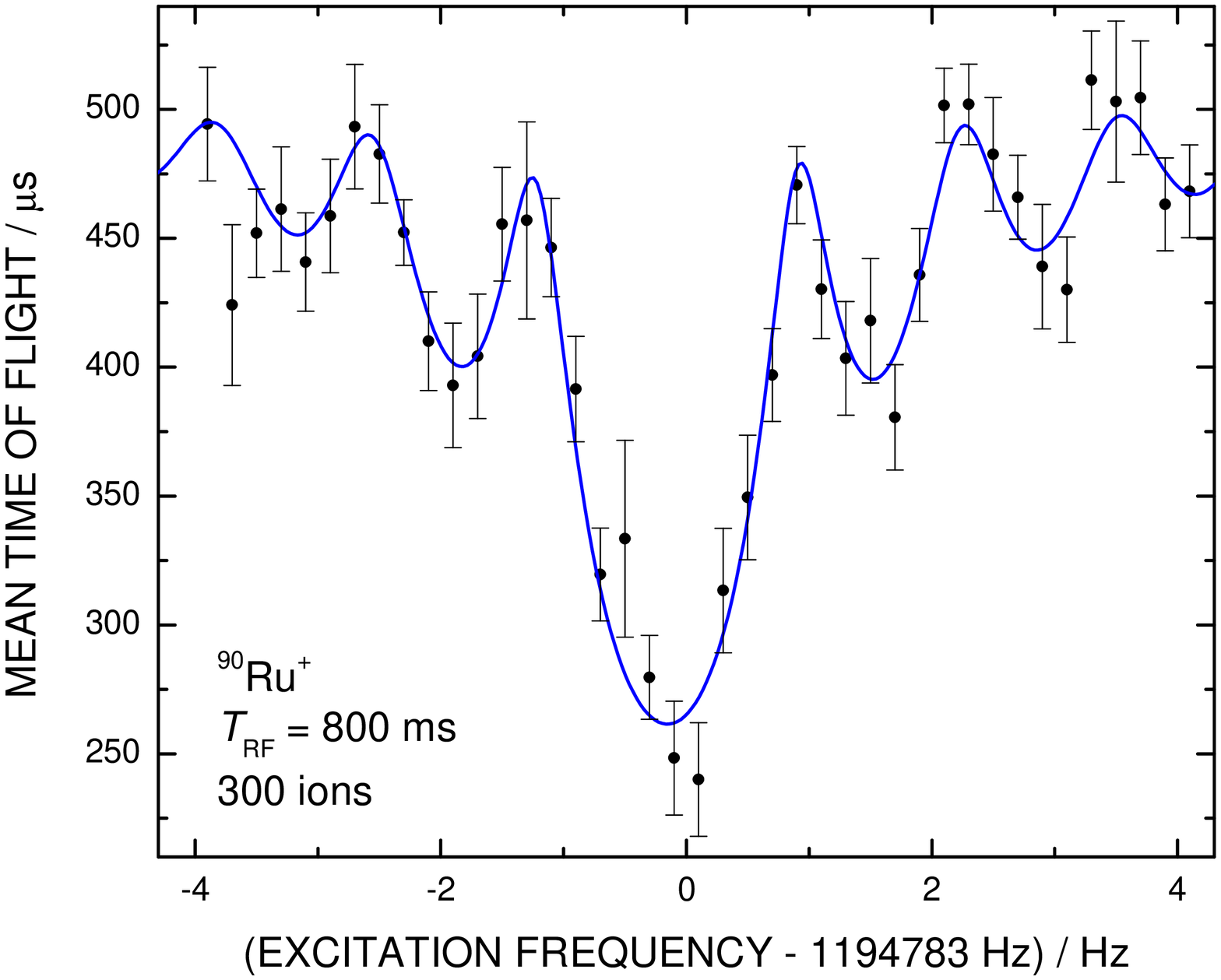}
\caption{\label{CRES_JYFL}Cyclotron resonance curves for
$^{91}\mbox{Tc}^{+}$ ions (top) measured at SHIPTRAP and $^{90}\mbox{Ru}^{+}$
ions (bottom) measured at JYFLTRAP. The duration of the RF excitation 
$T_{\text{RF}}$ and the number of detected ions are given.}
\end{figure}
In this way, individual isobars, or for some cases even isomers, can be selected. In the second Penning 
trap the mass $m$ of a stored ion species with charge $q$ is determined via a measurement of its cyclotron 
frequency $\nu_{\mbox{\scriptsize c}} = qB/(2 \pi m)$. A cyclotron resonance curve is obtained by the 
time-of-flight detection method: an increase in radial kinetic energy resulting 
from the resonant excitation of both radial ion motions with an RF-field at $\nu_{\mbox{\scriptsize c}}$
is detected by a reduction in the time of flight of the ejected ions towards a detector \cite{Graf1980,Koni1995a}. 
Figure \ref{CRES_JYFL} shows examples of cyclotron resonance curves for singly-charged $^{91}\mbox{Tc}$ and
$^{90}\mbox{Ru}$ ions from the SHIPTRAP and JYFLTRAP experiments, respectively.\\
Figure \ref{chart_JYFL_SHIP} displays a section of the nuclear chart below the doubly-magic shell closure of 
$^{100}\mbox{Sn}$. The nuclides studied at the SHIPTRAP (dark grey) and JYFLTRAP (light grey) facilities 
are approaching the $N = Z$ line. A possible pathway of the rp-process for steady-state burning (from Schatz \cite{Scha2001}, 
solid lines) is shown together with a possible path of the $\nu$p-process (this work, dashed lines). In Table 
\ref{tab:NUBASE}, half-lives $T_{1/2}$, spin-parities $I^{\pi}$, and excitation energies of the
first isomeric states $E_{\text{ex}}$ are listed for the studied nuclides according to the latest
NUBASE compilation \cite{NUBA2003}. In the calculation of reaction networks for astrophysical purposes, a preceding 
unambiguous mass-to-state assignment is essential. This overview helps to judge if the presence of a particular state 
can be excluded right away during our measurement process: states with a half-life $T_{1/2} \le
10~\mbox{ms}$ will not reach the setup. On the other hand, a separation of isomers will not be 
feasible if the excitation energy $E_{\text{ex}}$ of the specific ion is smaller than the mass resolution \footnote{This quantity depends on the cyclotron frequency $\nu_{\mbox{\scriptsize c}}$ and the duration $T_{\text{RF}}$ of the measurement.}; for example, $E_{\text{ex}} \approx 75~\mbox{keV}$ was employed for $^{88}\mbox{Tc}$. The mass-to-state assignment in this work is based on the present knowledge on these nuclides 
and is discussed in Sec. \ref{disc}.\\
\begin{center}
\begin{table}[here]
\caption{\label{tab:NUBASE}Properties of the nuclides investigated in this
work with half-life $T_{1/2}$, spin-parity $I^{\pi}$, and
excitation energy of the first isomeric states $E_{\text{ex}}$
from Ref. \cite{NUBA2003}. `*' denotes nuclides where the uncertainty
of the excitation energy $E_{\text{ex}}$ is larger than half of
the energy ($\sigma \ge E_{\text{ex}} / 2$) and `\&' those cases
where the ordering of the ground and isomeric states have been
reversed compared to the Evaluated Nuclear Structure Data
File (ENSDF) \cite{ENSD2008}. The parenthesis in the third column indicates uncertain
values of spin and/or parity, while `\#' indicates a value that is
estimated from systematic trends from neighboring nuclides with the 
same $Z$ and $N$ parities. }
\begin{ruledtabular}
\begin{tabular}{lrcccc}
Nuclide & &\multicolumn{2}{c}{Half-Life} & Spin/Parity  & Excitation\\
        & &\multicolumn{2}{c}{$T_{1/2}$} & $I^{\pi}$    & energy $E_{\text{ex}}$ / keV\\
\hline
$^{84}\mbox{Y}$     &   *   &   4.6 &   s     &   $1^{+}$         &                \\
$^{84}\mbox{Y}^{m}$ &   *   &   39.5&   min   &   ($5^{-}$)       &  -80(190)      \\
$^{87}\mbox{Zr}$    &       &   1.68&   h     &   $(9/2)^{+}$     &                \\
$^{87}\mbox{Zr}^{m}$&       &   14.0&   s     &   $(1/2)^{-}$     & 335.84(19)     \\
$^{88}\mbox{Mo}$    &       &   8.0 &   min   &   $0^{+}$         &                \\
$^{89}\mbox{Mo}$    &       &   2.11&   min   &   ($9/2^{+}$)     &                \\
$^{89}\mbox{Mo}^{m}$&       &   190 &   ms    &   ($1/2^{-}$)     & 387.5(2)       \\
$^{88}\mbox{Tc}$    &   *   &   5.8 &   s     &   (2,3)           &                \\
$^{88}\mbox{Tc}^{m}$&   *   &   6.4 &   s     &   (6,7,8)         &   0(300)\#     \\
$^{89}\mbox{Tc}$    &       &   12.8&   s     &   ($9/2^{+}$)     &                \\
$^{89}\mbox{Tc}^{m}$&       &   12.9&   s     &   ($1/2^{-}$)     &  62.6(5)       \\
$^{90}\mbox{Tc}$    &   *\& &   8.7 &   s     &   $1^{+}$         &                \\
$^{90}\mbox{Tc}^{m}$&   *\& &   49.2&   s     &   ($8^{+}$)       &  310(390)      \\
$^{91}\mbox{Tc}$    &       &   3.14&   min   &   $(9/2)^{+}$     &                \\
$^{91}\mbox{Tc}^{m}$&       &   3.3 &   min   &   $(1/2)^{-}$     &   139.3(3)     \\
$^{92}\mbox{Tc}$    &       &   4.25&   min   &   $(8)^{+}$       &                \\
$^{92}\mbox{Tc}^{m}$&       &   1.03&   $\mu$s&   $(4^{+})$       & 270.15(11)     \\
$^{90}\mbox{Ru}$    &       &   11  &   s     &   $0^{+}$         &                \\
$^{91}\mbox{Ru}$    &   *   &   9   &   s     &   ($9/2^{+}$)     &                \\
$^{91}\mbox{Ru}^{m}$&   *   &   7.6 &   s     &   ($1/2^{-}$)     &    80(300)\#   \\
$^{92}\mbox{Ru}$    &       &   3.65&   min   &   $0^{+}$         &                \\
$^{93}\mbox{Ru}$    &       &   59.7&   s     &   $(9/2)^{+}$     &                \\
$^{93}\mbox{Ru}^{m}$&       &   10.8&   s     &   $(1/2)^{-}$     &  734.4(1)      \\
$^{93}\mbox{Ru}^{n}$&       &   2.20&   $\mu$s&   $(21/2)^{+}$    &  2082.6(9)     \\
$^{94}\mbox{Ru}$    &       &   51.8&   min   &   $0^{+}$         &                \\
$^{94}\mbox{Ru}^{m}$&       &   71  &   $\mu$s&   ($8^{+}$)       & 2644.55(25)    \\
$^{92}\mbox{Rh}$    &       &   4.3 &   s     &   ($6^{+}$)\footnotemark[1]       &   \\
$^{93}\mbox{Rh}$    &       &   13.9&   s     &   $9/2^{+}$\#     &                \\
$^{94}\mbox{Rh}$    &   *   &   70.6&   s     &   ($2^{+},4^{+}$) &                \\
$^{94}\mbox{Rh}^{m}$&   *   &   25.8&   s     &   ($8^{+}$)       & 300(200)\#     \\
$^{95}\mbox{Rh}$    &       &   5.02&   min   &   $(9/2)^{+}$     &                \\
$^{95}\mbox{Rh}^{m}$&       &   1.96&   min   &   $(1/2)^{-}$     & 543.3(3)       \\
$^{94}\mbox{Pd}$    &       &   9.0 &   s     &   $0^{+}$         &                \\
$^{94}\mbox{Pd}^{m}$&       &   530 &   ns    &   ($14^{+}$)      & 4884.4(5)      \\
$^{95}\mbox{Pd}$    &       &   10\#&   s     &   $9/2^{+}$\#     &                \\
$^{95}\mbox{Pd}^{m}$&       &   13.3&   s     &   ($21/2^{+}$)    &   1860(500)\#\footnotemark[2]  \\
$^{96}\mbox{Pd}$    &       &   122 &   s     &   $0^{+}$         &                \\
$^{96}\mbox{Pd}^{m}$&       &   1.81&   $\mu$s&   $8^{+}$         & 2530.8(1)      \\
\end{tabular}
\end{ruledtabular}
\footnotemark[1]{A low-spin isomeric state with $T_{1/2} = 0.53~\mbox{s}$ has been observed in Ref. \cite{Dean2004}.}
\footnotetext[2]{An excitation energy of about $2~\mbox{MeV}$ was obtained in Ref. \cite{Nolt1982}.}
\end{table}
\end{center}
%%
\clearpage
%
%%%%%%%%%%%%%%%%%%%%%%%%%%%%%%%%%%%%%%%%%%%%%%%%%%%%%%%%%%%%%%%%%%%%%%%%%%%%%%%%%%%%%%%%%%%%%%%%%
%%%%%%%%%%%%%%%%%%%%%%%%%%%%%%%%%%%%%%%%%%%%%%%%%%%%%%%%%%%%%%%%%%%%%%%%%%%%%%%%%%%%%%%%%ANALYSIS
%%%%%%%%%%%%%%%%%%%%%%%%%%%%%%%%%%%%%%%%%%%%%%%%%%%%%%%%%%%%%%%%%%%%%%%%%%%%%%%%%%%%%%%%%%%%%%%%%%
\section{\label{ana}Data analysis}
The results of our mass measurements are given as frequency ratios, 
$r=\nu_{\textrm{c,ref}}/\nu_{\textrm{c}}$, between the cyclotron frequencies of a reference 
ion and the ion of interest. This ratio can be written as
\begin{equation}
r=\frac{\nu_{\textrm{c,ref}}}{\nu_{c}}=\frac{z_{\textrm{ref}}}{z}\cdot\frac{m-(z
\cdot m_{e})}{m_{\textrm{ref}}-(z_{\textrm{ref}}\cdot m_{e})}\ .
\end{equation}
In order to determine the mass $m$ from this expression one has to consider the inevitable 
drift of the magnetic field strength $B$, which has to be known at the moment when the ion 
of interest was studied. To this end, the measurement of a reference ion with well-known mass, 
before and after the measurement of the ion of interest, is used to determine the actual value of 
\textit{B}. In the case of singly-charged ions $z=z_{\textrm{ref}}=1$ the mass value is
calculated by
\begin{equation}\label{mass}
    m=r (m_{\textrm{ref}}-m_{e})+m_{e}\, ,
\end{equation}
where $m_{e}$ is the electron mass. The physics of a stored ion in a Penning trap including its motion as well as 
influences from field imperfections are extensively described in Refs. \cite{Brow1982,Brow1986,Boll1990,Kret1992a,Kret1992b}. 
Contributions to the total uncertainty in a 
frequency determination were quantified in systematic experimental studies at the ISOLTRAP mass spectrometer \cite{Blau2003,Kell2003} 
and the analysis of the data presented here was followed accordingly. In the following, a concise description of 
the contributions to the total uncertainty is given.

\begin{enumerate}
\item A least-squares fit of the theoretical line shape \cite{Koni1995a} to the experimental data 
points was initially performed. From this fit the cyclotron frequency $\nu_{c}$ and its 
statistical uncertainty $\sigma_{s}$ were obtained. The latter is inversely proportional to the number 
of detected ions, {\it i.e.} $\sigma_{s} \propto 1/\sqrt{N_{\text {ions}}}$. 
\item A count-rate class analysis \cite{Kell2003} was performed to correct possible frequency shifts 
from influences of contaminants, as, for example, other isobars or isomeric states. In some cases 
such an analysis was not feasible due to insufficient statistics. Here, the uncertainty of the cyclotron 
resonance frequency was increased according to a comparison between results without and with count-rate class 
analysis in high-statistics measurements of other short-lived nuclides. This accounts for 
the larger uncertainty after such an analysis, in order not to even favor those measurements 
with particularly low statistics. 
\item A pair of measurements of the reference ion, before and after the
measurement of the investigated radioactive ion, is performed to
calibrate the magnetic field \textit{B}. A linear interpolation is
used to calculate the value of $B$ at the time of the measurement
of the radioactive ion of interest. With these data a frequency
ratio $r=\nu_{\textrm{c,ref}}/\nu_{\textrm{c}}$ was calculated.
Fluctuations of the magnetic field strength in addition to the
linear trend are assumed to be of statistical nature. The longer the duration
between reference measurements, the larger the spread of such
fluctuations. Therefore this time-dependent uncertainty is
quadratically added to the uncertainty of each frequency ratio.
Their magnitudes were determined in long-term frequency
measurements of ions, $^{133}\mbox{Cs}^{+}$ and $^{85}\mbox{Rb}^{+}$ for SHIPTRAP
\cite{Raut2006}, and $^{54}\mbox{Fe}^{+}$ for JYFLTRAP
\cite{Raha2007}. Accordingly, the time-dependent quantities of
$\sigma_{B}(\nu_{\text{ref}})/\nu_{\text{ref}}=4.4\cdot10^{-9}\,\mbox{h}^{-1}\cdot\Delta t$ and $\sigma_{B}(\nu_{\text{ref}})/\nu_{\text{ref}}= %3.22\cdot10^{-11}\,\mbox{min}^{-1}\cdot\Delta t~(=
 1.9\cdot10^{-9}\,\mbox{h}^{-1}\cdot\Delta t$ 
have been used in these data, respectively.
\item The weighted mean of the available frequency ratios for a
given nuclide was calculated and the consistency of internal
versus external uncertainties was compared \cite{Birg1932}. The ratios of both were 
found to be very close to unity in all cases for both sets of data
indicating that only statistical fluctuations due to random 
deviations are present in the experimental data. The
larger value of the two was taken as the final uncertainty. In
this way the uncertainties are always given conservatively.

\item By means of cross-reference mass measurements with carbon
clusters \cite{Chau2007}, which provide absolute frequency ratios, a relative
mass-dependent shift of $\sigma_{m}(r)/r=(1.1\pm 1.7)\cdot 10^{-10}
/u\cdot \Delta m$ was obtained at SHIPTRAP, where
$\Delta m$ indicates the difference in mass between the
radionuclide of interest and the reference ion $(m -
m_\text{ref})$. At the JYFLTRAP
facility this quantity has been determined via a mass comparison
of oxygen $^{16}\mbox{O}_2^{+}$ to xenon $^{132}\mbox{Xe}^{+}$
ions to be $|\sigma_{m}(r)/r|=7\cdot 10^{-10} /u\cdot \Delta m$. 
For the SHIPTRAP data, the averaged frequency
ratios were corrected by this mass-dependent frequency shift and
the same quantity was quadratically added to its uncertainty,
whereas for JYFLTRAP no correction was applied due to the
undetermined sign of this quantity. Since the accurate frequency ratios 
of carbon clusters of different sizes are known without any uncertainty 
\footnote{The estimated values of the molecular binding energy/atom of the
carbon clusters varies from 3.1 eV in $^{12}\mbox{C}_{2}$ to 7.0
eV in $^{12}\mbox{C}_{60}$ \cite{Toma1991}, therefore it is
negligible at the present uncertainty level for mass measurements
at both facilities.} 
the relative residual uncertainty was determined to be 
$\sigma_{\text{res}}(r)/r=4.5\cdot10^{-8}$ with carbon 
clusters at SHIPTRAP \cite{Chau2007}. Since this quantity 
has not yet been experimentally determined at JYFLTRAP, the 
deviation between previous JYFLTRAP data and precise experimental AME2003 
values in this mass region have been compared. Their relative average
deviation is about $5\cdot10^{-8}$. The residual uncertainty and the latter 
estimate have been added to give the final uncertainty of the frequency 
ratio in either experiment.
\end{enumerate}
\section{\label{res}Results}
\begin{figure*}[h]
\includegraphics[width=0.7\textwidth]{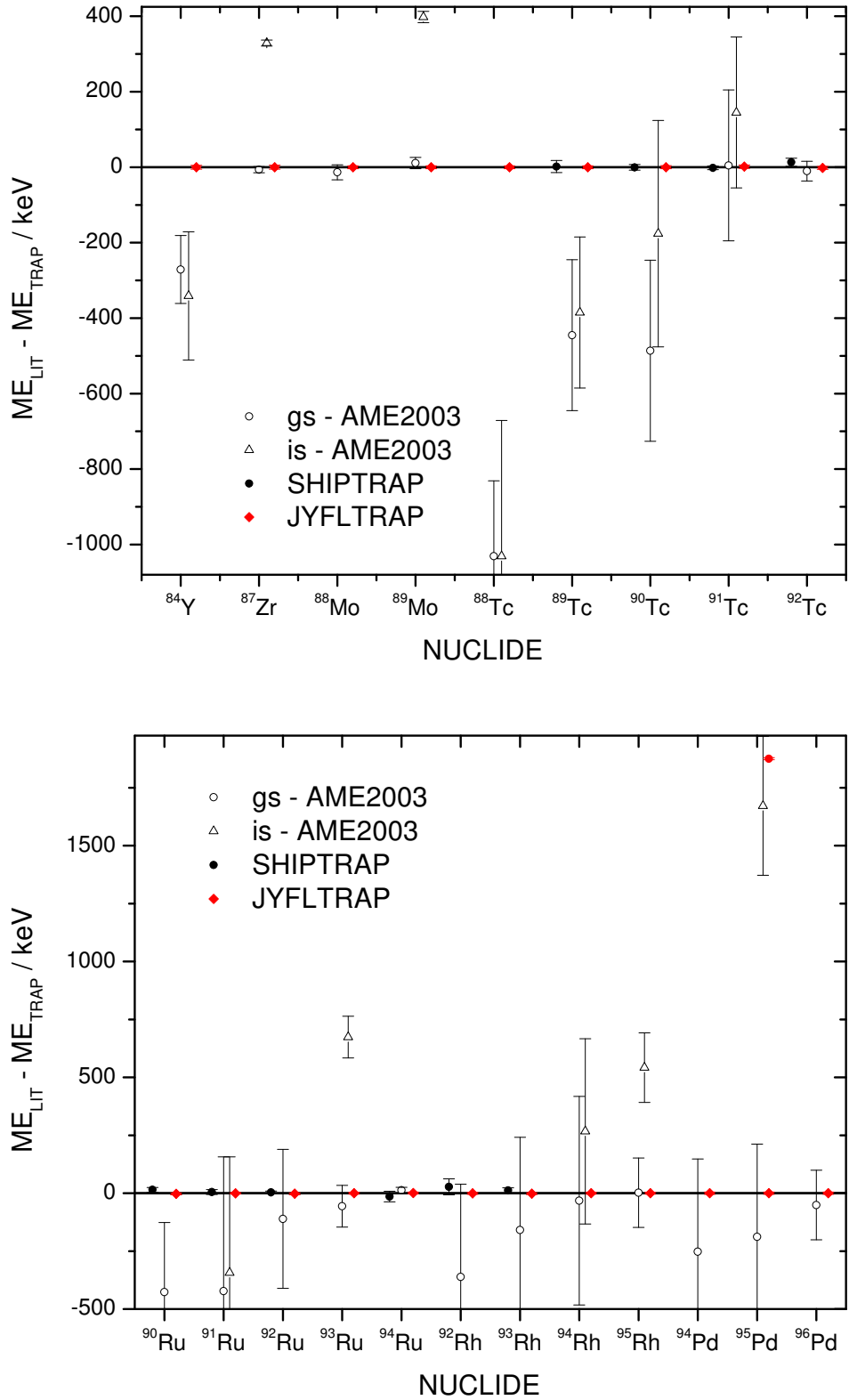}
\caption{\label{Ru to Pd}Differences between experimental mass
excess values from SHIPTRAP and JYFLTRAP with respect to the
literature (AME2003) \cite{NUBA2003,Audi2003a}. When data from
both trap experiments are available, the weighted average is taken
as the final result $ME_{\text{TRAP}}$ and set to zero. Open
triangles indicate the position of the first excited
isomeric state with respect to the AME2003 ground-state mass value. Color figure online.}
\end{figure*}

The data reported here were obtained during several on-line runs at the
SHIPTRAP/GSI and JYFLTRAP/IGISOL facilities. The measurements at
SHIPTRAP were performed in July 2006 \cite{Ferr2007}. In that work, the UNILAC primary
beam of $^{40}\mbox{Ca}$ ions at $200~\mbox{MeV}$ energy was used in
combination with a $0.5~\mbox{mg/cm}^{2}$ target of
$^{58}\mbox{Ni}$ (enriched to $99.9\%$) in order to produce
neutron-deficient species of technetium, ruthenium, and rhodium in
the region around $A = 90$ via fusion-evaporation reactions. The
identical reaction was used in August 2006 at JYFLTRAP \cite{Elom2009}, where a 
$^{\text{nat}}\mbox{Ni}$ target
\footnote{$^{58}\mbox{Ni}:68.0769(89)\%$, $^{60}\mbox{Ni}:26.2231(77)\%$,\\ 
$^{61}\mbox{Ni}:1.1399(6)\%$, $^{62}\mbox{Ni}:3.6345(17)\%$,\\
$^{64}\mbox{Ni}:0.9256(9)\%$, http://www.webelements.com.}
of $3.6~\mbox{mg/cm}^2$ foil thickness
was used in combination with a beam of $189~\mbox{MeV}$ (JYFLTRAP II) or $220~\mbox{MeV}$ 
(JYFLTRAP III) $^{40}\mbox{Ca}$ ions. In addition, a beam of $^{36}\mbox{Ar}$ at $222~\mbox{MeV}$ 
(JYFLTRAP I) energy was employed to produce $^{84}\mbox{Y}$,
$^{87}\mbox{Zr}$, $^{89}\mbox{Mo}$, and $^{88}\mbox{Tc}$ ions. In a second experiment 
with JYFLTRAP in December 2006 a $^{40}\mbox{Ca}$ beam at $205~\mbox{MeV}$ (JYFLTRAP IV) and
$170~\mbox{MeV}$ energy (JYFLTRAP V) was used to produce the $A = 91$ isobars of technetium and ruthenium and the 
isotopes of palladium, respectively.\\
All results from the cyclotron frequency determination at JYFLTRAP
and SHIPTRAP are presented in Table \ref{tab:results}. In the
second column the experiment is indicated to refer to the
parameters for the production of these exotic nuclides as given above. The third column
lists the number of detected ions in all cyclotron resonance curves and in
column four and five the reference nuclide and the resulting
frequency ratios $r=\nu_{\textrm{c,ref}}/\nu_{\textrm{c}}$ are
given. For ten nuclides the cyclotron frequencies have been
determined at both experiments, SHIPTRAP and JYFLTRAP. In these cases, a
weighted mean of the frequency ratio is given. From the
experimental results the mass excess of the nuclide under study is
derived using the atomic masses of $^{97}\mbox{Mo}$,
$^{16}\mbox{O}$, $^{85}\mbox{Rb}$, or $^{94}\mbox{Mo}$ from the
Atomic Mass Evaluation, AME2003 \cite{Audi2003a}. The final values
are compared with those from AME2003 and their differences are given in 
the last column.\\
A graphical representation of the mass differences between AME2003 and this work is shown in Fig. 
\ref{Ru to Pd}. For the ten nuclides studied by both experiments, the Penning trap results 
agree in all cases perfectly with each other. They yield final weighted averages, which, along with 
the individual JYFLTRAP data, are set to zero in the diagram. The literature values are
indicated by open symbols, whereas the positions of excited isomeric nuclear states are depicted as
triangles. Within each of the individual isotopes an increasing deviation is visible departing from 
the valley of stability. This might be partly explained due to literature values having been extrapolated 
from systematic trends (see Table \ref{tab:results}). Note that for $^{95}\mbox{Pd}$ the mass 
values of the ground as well as of the ($21/2^{+}$) isomeric state were determined. 
\begin{table*}
\caption{\label{tab:results}Results from cyclotron frequency determinations at SHIPTRAP and JYFLTRAP. The experimental
parameters for the production of these nuclides are given in the text. The cyclotron frequency ratio $r$ to a singly-charged reference ion
`Ref' is given in the fifth column. The data from both experiments are averaged for the frequency ratios $r$ or the deduced mass excess
values $ME$. The latter were derived using the atomic masses of $^{97}\mbox{Mo}$, $^{16}\mbox{O}$, $^{85}\mbox{Rb}$, or $^{94}\mbox{Mo}$ 
from the Atomic Mass Evaluation 2003 \cite{Audi2003a}. The final values are compared with those from AME2003 and their differences 
$\Delta$ are given in the last column. `\#' indicates a value that is estimated from systematic experimental trends.}
\begin{ruledtabular}
\begin{tabular}{lcrclllc}
Nuclide             & Experiment & $N_{\text{ions}}$ & Ref
&Frequency ratio $r$ & $ME_{\text{TRAP}}$ / keV &
$ME_{\text{AME2003}}$ / keV & $\Delta_{\text{AME2003-TRAP}}$ /
keV\\
\hline\\
$^{84}\mbox{Y}$     &   JYFLTRAP I     &   1918&   $^{97}\mbox{Mo}$    &   1.031056762(53)\footnotemark[1]& -73922(19)\footnotemark[2] & -74160(90)     &   -238(92) \\
%$^{84}\mbox{Y}^{m}$ &   JYFL I     &       &   $^{97}\mbox{Mo}$    &                     &                   &   -74230(170)    &   -341(170)\\
$^{87}\mbox{Zr}$    &   JYFLTRAP I     &   3673&   $^{97}\mbox{Mo}$    &   1.061954361(54)\footnotemark[3]&-79341.4(5.3)   &   -79348(8)      &   -7(10)   \\
%$^{87}\mbox{Zr}^{m}$&   JYFL I     &       &   $^{97}\mbox{Mo}$    &                     &                   &   -79012(8)      &   329(10)  \\
$^{88}\mbox{Mo}$    &   JYFLTRAP II    &   8576&   $^{85}\mbox{Rb}$    &   1.035450878(48)   &   -72686.5(3.8)   &   -72700(20)     &   -14(20)  \\
$^{89}\mbox{Mo}$    &   JYFLTRAP I     &   2871&   $^{85}\mbox{Rb}$    &   1.047198443(49)   &   -75015.0(3.9)   &   -75004(15)     &   11(15)   \\
%$^{89}\mbox{Mo}^{m}$&   JYFL I     &       &   $^{85}\mbox{Rb}$    &                     &                   &   -74617(15)     &   398(15)  \\
$^{88}\mbox{Tc}$    &   JYFLTRAP I     &   8715&   $^{85}\mbox{Rb}$    &   1.035590047(48)   &   -61679(87)\footnotemark[4]   &   -62710(200)\#  &   -1031(218)\#\\
%$^{88}\mbox{Tc}^{m}$&   JYFL I     &       &   $^{85}\mbox{Rb}$    &                     &                   &   -62710(360)\#  &   -1031(360)\\
$^{89}\mbox{Tc}$    &   JYFLTRAP III   &   9817&   $^{85}\mbox{Rb}$    &   1.047294785(48)   & \\%  -67394.8(3.8)   &   -67840(200)\#  &   -445(200) \\
$^{89}\mbox{Tc}$    &   SHIPTRAP       &   879 &   $^{85}\mbox{Rb}$    &   1.047294809(203)  & \\%  -67393.0(16.0)  &   -67840(200)\#  &   -447(201) \\
$^{89}\mbox{Tc}$    &   AVERAGE        &       &   $^{85}\mbox{Rb}$    &   1.047294786(47)   &   -67394.8(3.7)   &   -67840(200)\#  &   -445(200)\# \\
%$^{89}\mbox{Tc}^{m}$&   AVERAGE        &       &   $^{85}\mbox{Rb}$    &                     &                   &   -67780(200)\#  &   -385(200) \\
$^{90}\mbox{Tc}$&   JYFLTRAP III   &   8503&   $^{85}\mbox{Rb}$    &   1.059029703(49)   & \\%  -70723.6(3.9)   &   -71210(240)    &   -486(240) \\
$^{90}\mbox{Tc}$&   SHIPTRAP       &   2516&   $^{85}\mbox{Rb}$    &   1.059029697(94)   & \\%  -70724.1(7.5)   &   -71210(240)    &   -486(240) \\
$^{90}\mbox{Tc}$&   AVERAGE        &       &   $^{85}\mbox{Rb}$    &   1.059029702(43)   &   -70723.7(3.4)   &   -71210(240)    &   -486(240) \\
%$^{90}\mbox{Tc}^{m}$&   AVERAGE        &       &   $^{85}\mbox{Rb}$    &                     &                   &   -70900(300)    &   -176(300) \\
$^{91}\mbox{Tc}$    &   JYFLTRAP IV    &   2291&   $^{94}\mbox{Mo}$    &   0.968194724(47)   &   -75983.4(4.5) \\%  &   -75980(200)    &   3(200)    \\
$^{91}\mbox{Tc}$    &   SHIPTRAP       &   5890&   $^{85}\mbox{Rb}$    &   1.070740167(63)   &   -75986.5(5.0) \\%  &   -75980(200)    &   7(200)    \\
$^{91}\mbox{Tc}$    &   AVERAGE        &       &                       &                     &   -75984.8(3.3)   &   -75980(200)    &   5(200)    \\
%                    &   AVERAGE        &       &                       &                     &                   &   -75840(200     &   145(200)  \\
$^{92}\mbox{Tc}$    &   JYFLTRAP II    &   9399&   $^{85}\mbox{Rb}$    &   1.082480001(50)   & \\%  -78926.4(3.9)   &   -78935(26)     &   -9(26)    \\
$^{92}\mbox{Tc}$    &   SHIPTRAP       &   7638&   $^{85}\mbox{Rb}$    &   1.082480188(137)  & \\%  -78911.6(10.8)  &   -78935(26)     &   -23(28)   \\
$^{92}\mbox{Tc}$    &   AVERAGE        &       &   $^{85}\mbox{Rb}$    &   1.082480023(47)   &   -78924.7(3.7)   &   -78935(26)     &   -10(26)   \\
$^{90}\mbox{Ru}$    &   JYFLTRAP III   &   823 &   $^{85}\mbox{Rb}$    &   1.059103508(55)   & \\%  -64886.1(4.4)   &   -65310(300)\#  &   -424(300)  \\
$^{90}\mbox{Ru}$    &   SHIPTRAP       &   259 &   $^{85}\mbox{Rb}$    &   1.059103727(127)  & \\%  -64868.8(10.0)  &   -65310(300)\#  &   -441(300)  \\
$^{90}\mbox{Ru}$    &   AVERAGE        &       &   $^{85}\mbox{Rb}$    &   1.059103543(50)   &   -64883.3(4.0)   &   -65310(300)\#  &   -427(300)\#  \\
$^{91}\mbox{Ru}$    &   JYFLTRAP II    &   7279&   $^{85}\mbox{Rb}$    &   1.070838119(49)   &   -68239.1(3.9)  \\% &   -68660(580)\#  &   -421(580)  \\
$^{91}\mbox{Ru}$    &   JYFLTRAP IV    &   1510&   $^{94}\mbox{Mo}$    &   0.968283302(50)   &   -68235.3(4.7)  \\% &   -68660(580)\#  &   -425(580)  \\
$^{91}\mbox{Ru}$    &   SHIPTRAP       &   1111&   $^{85}\mbox{Rb}$    &   1.070838208(131)  &   -68232.0(10.4) \\% &   -68660(580)\#  &   -428(580)  \\
$^{91}\mbox{Ru}$    &   AVERAGE        &       &                       &                     &   -68237.1(2.9)  &   -68660(580)\#  &   -423(580)\#  \\
%$^{91}\mbox{Ru}^{m}$&   AVERAGE        &       &                       &                     &                   &   -68580(500)    &   -343(500)  \\
$^{92}\mbox{Ru}$    &   JYFLTRAP III   &   3744&   $^{85}\mbox{Rb}$    &   1.082538479(51)   & \\%  -74301.1(4.0)   &   -74410(300)\#  &   -109(300)  \\
$^{92}\mbox{Ru}$    &   SHIPTRAP       &   5691&   $^{85}\mbox{Rb}$    &   1.082538554(67)   & \\%  -74295.2(5.3)   &   -74410(300)\#  &   -115(300)  \\
$^{92}\mbox{Ru}$    &   AVERAGE        &       &   $^{85}\mbox{Rb}$    &   1.082538507(41)   &   -74299.0(3.2)   &   -74410(300)\#  &   -111(300)\#  \\
$^{93}\mbox{Ru}$    &   JYFLTRAP III   &   5748&   $^{85}\mbox{Rb}$    &   1.094278655(51)   &   -77214.0(4.0)   &   -77270(90)     &   -56(90)    \\
%$^{93}\mbox{Ru}^{m}$&   JYFL III   &       &   $^{85}\mbox{Rb}$    &                     &                   &   -76540(90)     &   674(90)    \\
$^{94}\mbox{Ru}$    &   JYFLTRAP II    &   5589&   $^{85}\mbox{Rb}$    &   1.105987814(53)   & \\%  -82580.2(4.2)   &   -82568(13)     &   12(14)     \\
$^{94}\mbox{Ru}$    &   SHIPTRAP       &   1093&   $^{85}\mbox{Rb}$    &   1.105987624(289)  & \\%  -82595.1(22.8)  &   -82568(13)     &   27(26)     \\
$^{94}\mbox{Ru}$    &   AVERAGE        &       &   $^{85}\mbox{Rb}$    &   1.105987808(52)   &   -82580.6(4.1)   &   -82568(13)     &   13(14)     \\
$^{92}\mbox{Rh}$    &   JYFLTRAP III   &   1804&   $^{85}\mbox{Rb}$    &   1.082681373(55)   & \\%  -62999.1(4.3)   &   -63360(400)\#  &   -361(400)  \\
$^{92}\mbox{Rh}$    &   SHIPTRAP       &   220 &   $^{85}\mbox{Rb}$    &   1.082681725(434)  & \\%  -62971.2(34.4)  &   -63360(400)\#  &   -389(401)  \\
$^{92}\mbox{Rh}$    &   AVERAGE        &       &   $^{85}\mbox{Rb}$    &   1.082681379(55)   &   -62999(15)\footnotemark[5]   &   -63360(400)\#  &   -361(400)\#  \\
$^{93}\mbox{Rh}$    &   JYFLTRAP III   &   1437&   $^{85}\mbox{Rb}$    &   1.094382341(53)   & \\ % -69013.0(4.2)   &   -69170(400)\#  &   -157(400)  \\
$^{93}\mbox{Rh}$    &   SHIPTRAP       &   1035&   $^{85}\mbox{Rb}$    &   1.094382514(141)  & \\ %  -68999.4(11.1)  &   -69170(400)\#  &   -171(400)  \\
$^{93}\mbox{Rh}$    &   AVERAGE        &       &   $^{85}\mbox{Rb}$    &   1.094382362(50)   &   -69011.3(3.9)   &   -69170(400)\#  &   -159(400)\#  \\
%$^{94}\mbox{Rh}^{m}$&   JYFL II,III&   1515&  $^{85}\mbox{Rb}$    &   1.106110107(54)   &   -72907.5(4.2)   &   -72640(400)  &   267(400)   \\
$^{94}\mbox{Rh}$     &   JYFLTRAP II,III&  1515&   $^{85}\mbox{Rb}$    &   1.106110107(54)   &   -72907.5(4.2)   &   -72940(450)\#  &   -33(450)\#   \\
$^{95}\mbox{Rh}$    &   JYFLTRAP II    &   1965&   $^{85}\mbox{Rb}$    &   1.117818397(53)   &   -78342.3(4.2)   &   -78340(150)    &   2(150)     \\
%$^{95}\mbox{Rh}^{m}$&   JYFL II    &       &   $^{85}\mbox{Rb}$    &                     &                   &   -77800(150)    &   542(150)   \\
$^{94}\mbox{Pd}$    &   JYFLTRAP V     &   1606&   $^{94}\mbox{Mo}$    &   1.000255075(49)   &   -66097.9(4.7)   &   -66350(400)\#  &   -252(400)\#  \\
$^{95}\mbox{Pd}$    &   JYFLTRAP V     &   1470&   $^{94}\mbox{Mo}$    &   1.010860017(50)   &   -69961.6(4.8)   &   -70150(400)\#  &   -188(400)\#  \\
$^{95}\mbox{Pd}^{m}$&   JYFLTRAP V     &   5307&   $^{94}\mbox{Mo}$    &   1.010881456(48)   &   -68086.2(4.7)   &   -68290(300)    &   -204(300)  \\
$^{96}\mbox{Pd}$    &   JYFLTRAP V     &   3570&   $^{94}\mbox{Mo}$    &   1.021438050(48)   &   -76179.0(4.7)   &   -76230(150)    &   -51(150)   \\
\end{tabular}
\end{ruledtabular}
\footnotetext[1]{This ratio was measured for $\mbox{YO}^{+}$
ions.}\footnotetext[2]{The original value of
$-73888.8(5.2)~\mbox{keV}$ was modified for an unknown mixture of
isomeric states.}\footnotetext[3]{This ratio was measured for
$\mbox{ZrO}^{+}$ ions.}\footnotetext[4]{The original error of 3.8 keV was increased due to the unknown level scheme of isomeric states.}
\footnotetext[5]{The original error of 4.3 keV was increased due to the unknown level scheme of isomeric states.}
\end{table*}
%\clearpage
%
\begin{figure*}[h]
\includegraphics[width=0.7\textwidth]{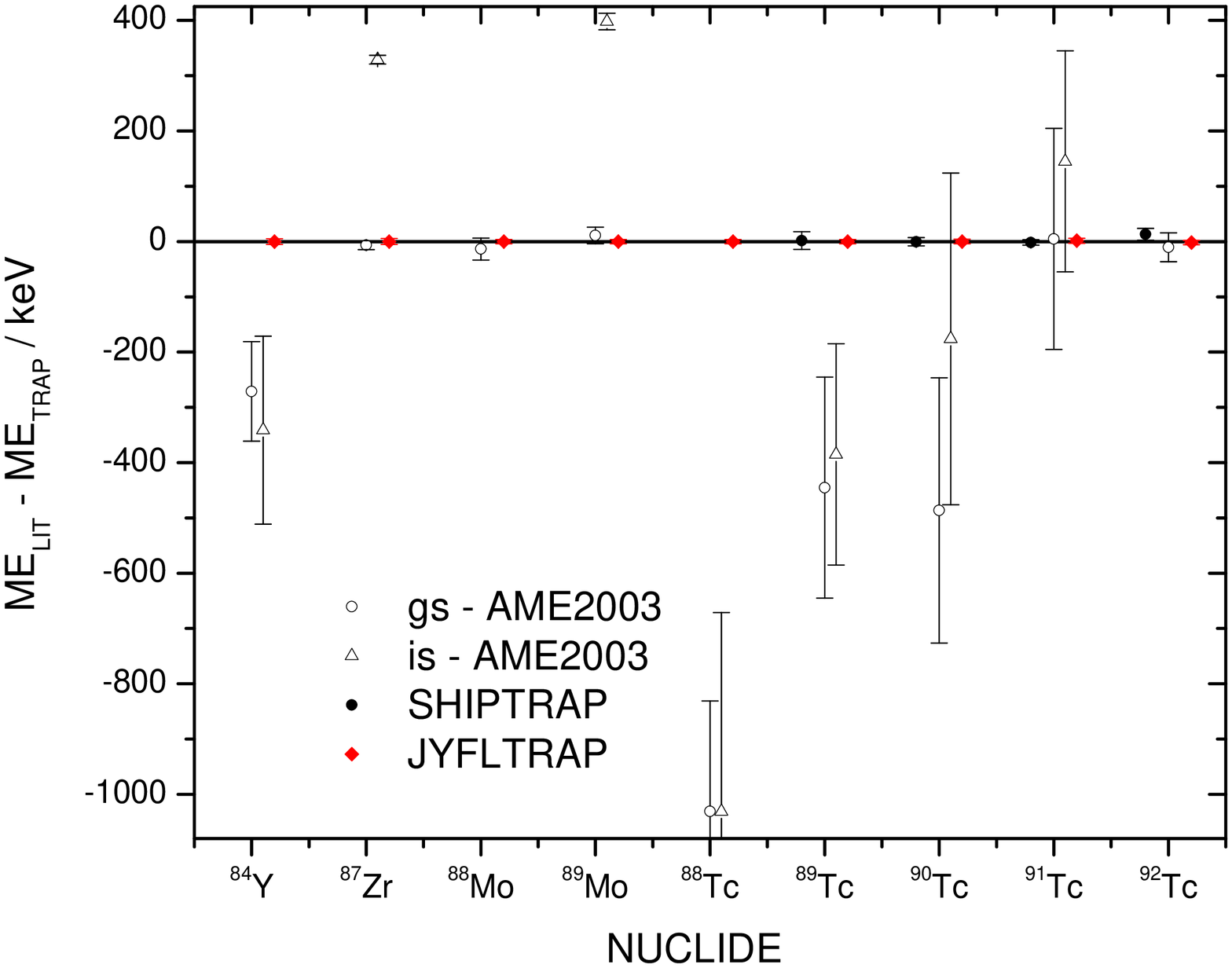}
%\caption{\label{Y to Tc}}
\end{figure*}
\clearpage
%%%%%%%%%%%%%%%%%%%%%%%%%%%%%%%%%%%%%%%%%%%%%%%%%%%%%%%%%%%%%%%%%%%%%%%%%%%%%%%%%%%%%%%%%%%%%%%%%%
%%%%%%%%%%%%%%%%%%%%%%%%%%%%%%%%%%%%%%%%%%%%%%%%%%%%%%%%%%%%%%%%%%%%%%%%%%%%%%%%%%%%%%%%DISCUSSION
%%%%%%%%%%%%%%%%%%%%%%%%%%%%%%%%%%%%%%%%%%%%%%%%%%%%%%%%%%%%%%%%%%%%%%%%%%%%%%%%%%%%%%%%%%%%%%%%%%
\section{\label{disc}Discussion}
In this discussion, some of the nuclides are grouped such as odd-even and even-odd 
ones with assigned spin values of $9/2^+$ and $1/2^-$, or nuclides where additional 
isomeric states are excluded due to their short half-lives. The remaining cases are discussed 
with increasing $Z$ and $A$. It should be noted that seven of the nuclides investigated in this 
work ($^{90-92}\mbox{Tc}$, $^{93,94}\mbox{Ru}$, and $^{94,95}\mbox{Rh}$), were already studied 
with the Canadian Penning Trap (CPT) at the Argonne National Laboratory \cite{Clar2005}. Since 
these data are published only in the form of a diagram, a quantitative 
comparison is omitted here. However, a qualitative comparison indicates similar mass values, 
obtained in both experiments following a production with heavy-ion fusion-evaporation reactions.\\

{\bf $^{84}\mbox{Y}$}: This nucleus has a 39.5-min isomer with
different assignments of spin values $I^{\pi} = 4^-$ \cite{Iafi1976},
$5^-$ \cite{List1981,Fire1996}, or $6^+$ \cite{Doer2000} and a
$I^{\pi} = 1^+$ low-spin state with a half-life of $4.6~\mbox{s}$.
The level scheme has been under continuous debate with claims that the
ground state is most likely the 39.5-min state \cite{Iafi1976}
as well as the shorter-lived $1^+$ state \cite{Fire1996}. The $Q_{\text{EC}}$
values of both $\beta$-decaying isomers \cite{List1981} place the
$5^-$ state energetically lower than the $1^+$ state although with a very high
uncertainty. Hence, their reversed placement according to the
systematics in AME2003 results in a negative value of the
excitation energy in Table \ref{tab:NUBASE}. The most recent
experiment, in-beam studies carried out at NBI (Copenhagen) and
Florida State University, propose a 40-min, $6^+$ ground state and
a $1^+$ isomer at an excitation energy of $67~\mbox{keV}$ \cite{Doer2000}.\\
However, the mass resolution employed in our study of around $100~\mbox{keV}$
for the measured $\mbox{YO}^{+}$ ion is not sufficient to resolve
ground and isomeric states and no count-rate class analysis has
been carried out due to low statistics. Therefore, a correction according to a treatment for an unknown
mixture of isomers \cite{Waps2003} has been applied, modifying the
original mass excess value from $-73888.8(5.2)~\mbox{keV}$ to $-73922(19)~\mbox{keV}$.\\

{\bf $^{88}\mbox{Mo}$}: The mass excess value determined at
JYFLTRAP agrees with the value of the AME2003,
which is deduced from the measured $Q$ value of the
$^{92}\mbox{Mo}(\alpha,^{8}\mbox{He})^{88}\mbox{Mo}$ reaction
studied at INS, Tokyo \cite{Kato1990}, having a precision of
$20~\mbox{keV}$.\\

\begin{figure}
\includegraphics[width=0.48\textwidth]{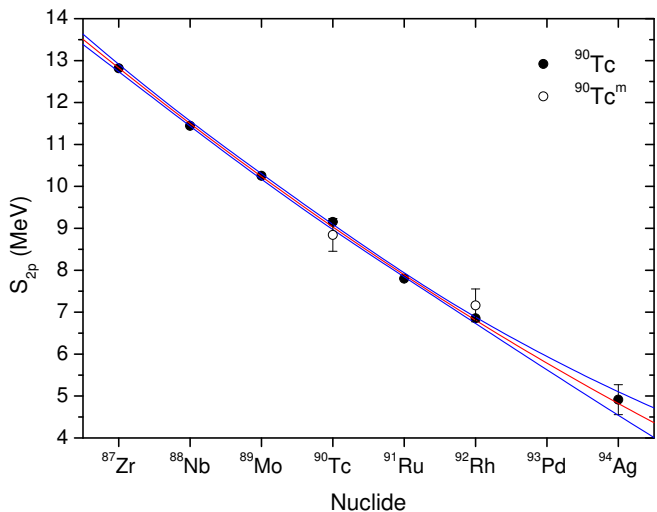}
\caption{\label{47EX}Plot of two-proton separation
energies for $N = 47$ isotones. The binding energies for
$^{85}\mbox{Sr}$ and $^{86}\mbox{Y}$ were taken from AME2003 and
the value for $^{94}\mbox{Ag}$ was derived from the Coulomb
displacement energy and the newly-determined mass of
$^{94}\mbox{Pd}$. $S_{\text{2p}}$ energies with our mass value assigned to the 
ground or isomeric state of $^{90}\mbox{Tc}$ are indicated. This
diagram is discussed in a dedicated publication on the decay modes in
$^{94}\mbox{Ag}$ \cite{Kank2008}.}
\end{figure}

%%%%%%%%%%%%%%%%%%%%%%%%%%%%%%%%%%%%%%%%%%%%%%%%%%%%%%%%%%%%%%%%%%%%%%%%%%%%%%%%%%%%%%%%%%%%%%%%%%
%%%%%%%%%%%%%%%%%%%%%%%%%%%%%%%%%%%%%%%%%%%%%%%%%%%%%%%%%%%%%%%%%%%%%%%%%%%%%%%%%%%%%%%%TECHNETIUM
%%%%%%%%%%%%%%%%%%%%%%%%%%%%%%%%%%%%%%%%%%%%%%%%%%%%%%%%%%%%%%%%%%%%%%%%%%%%%%%%%%%%%%%%%%%%%%%%%%
{\bf $^{90}\mbox{Tc}$}: Prior to the Penning trap measurements, the
radioactive nuclide $^{90}\mbox{Tc}$ was produced by a (p,3n)
reaction on isotopically enriched $^{92}\mbox{Mo}$ at
$43~\mbox{MeV}$ beam energy and studied by means of $\beta$-decay
spectroscopy in two independent experiments at Foster Radiation
Laboratory in Canada \cite{Iafi1974,Oxor1981}. Two long-lived
states were found in both experiments with (i) $T_{1/2} =
8.7~\mbox{s}$ and $I^{\pi} = 1^+$ and (ii) $T_{1/2} =
49.2~\mbox{s}$ and $I^{\pi} = 8^+$ \cite{Oxor1981}. Following
$Q$-value measurements in coincidence with $\gamma$ rays in
$^{90}\mbox{Mo}$ the shorter-lived $1^+$ component was assigned
to the ground state \cite{Oxor1981}. The excitation energy of the
49.2-s level was deduced in the AME2003 \cite{Audi2003a} employing the data from 
Iafigliola \cite{Iafi1974} and Oxorn \cite{Oxor1981}, with the earlier being a measurement without a 
$\beta$-$\gamma$ coincidence. Here, the branching ratio to the 948-keV ($2^+)$ level
in $^{90}\mbox{Mo}$ is 22\% in contrast to Oxorn and Mark \cite{Oxor1981}, 
which give log $ft$ values that are vice versa. With a branching of 78\% \cite{Oxor1981}, 
the excitation energy to be derived from these data is $124(390)~\mbox{keV}$ instead of
$310(390)~\mbox{keV}$ \cite{NUBA2003}.\\
In two other experiments, $^{90}\mbox{Tc}$ was produced
via the fusion-evaporation reaction $^{58}\mbox{Ni}(^{36}\mbox{Ar},3\mbox{pn})$ at
$149~\mbox{MeV}$ beam energy \cite{Rudo1993} and as a decay product of 
laser-ionized $^{90}\mbox{Ru}$ produced in 
$^{58}\mbox{Ni}(^{36}\mbox{Ar},2\mbox{p}2\mbox{n})^{90}\mbox{Ru}$ reactions at
$180~\mbox{MeV}$ beam energy at the LISOL facility \cite{Dean2004}. In the latter 
experiment only low-spin states of $^{90}\mbox{Tc}$ were populated in the 
$\beta$ decay of $^{90}\mbox{Ru}$ ($I^{\pi} = 0^+$), while in the earlier in-beam 
experiment no low-spin state could be observed. This behavior agrees with the 
expectation that in fusion-evaporation reactions states at high
spin and excitation energies up to several MeV are populated due
to their high geometrical cross sections \cite{Bass1980}. Thus, a
considerable amount of the flux populates high-spin
states.\\
Rudolph {\it et al.} 
\cite{Rudo1993} refer to this level as the ground state, supported by the 
observation from experimental spin assignments in the neighboring 
odd-odd nuclides
$^{88}\mbox{Nb}$ ($N = 47$), $^{90}\mbox{Nb}$ ($N = 49$), $^{92}\mbox{Tc}$ ($N = 49$), 
and the predictions from shell-model
calculations.\\ 
In Dean {\it et al.} \cite{Dean2004} the $\beta$ decay of $^{90}\mbox{Ru}$ was 
studied and the low-spin and high-spin level schemes from shell-model calculations 
were compared with experimental results. In conclusion, these calculations are also 
suggesting a high-spin ground state for $^{90}\mbox{Tc}$.\\
Our production mechanism via fusion-evaporation reactions
would also favor the production of the high-spin isomer. A
careful examination of the SHIPTRAP data using the count-rate
class analysis \cite{Kell2003} shows no evidence of a possible
contamination in the trap, even regarding a smaller excitation energy. 
For these reasons we clearly assign the measured mass value to the $8^+$ level.\\
Within the $N = 47$ chain several new Penning trap measurements
were conducted in this work for the isotones from $^{87}\mbox{Zr}$ 
($Z = 40$) to $^{92}\mbox{Rh}$ ($Z = 45$) (see Fig. \ref{chart_JYFL_SHIP}).
In the attempt to interpolate the mass value for $^{93}\mbox{Pd}$,
the two-proton separation energies derived from these data were
plotted. Figure \ref{47EX} shows a smooth systematic behavior after
the assignment of our mass value to the ground state of $^{90}\mbox{Tc}$ according to
the justification given in Ref. \cite{Rudo1993}. If our mass value were assigned to the 
excited isomeric state \cite{Audi2003a,Oxor1981}, the smooth trend is
interrupted. We therefore regard the first possibility as the more
probable one. However, considering the prevalent discrepancies among the available
literature this nuclide should be addressed in a future
measurement.\\

{\bf $^{88}\mbox{Tc}$}: This nuclide was first identified in
in-beam $\gamma$-ray studies after fusion-evaporation reactions. A
study of the $^{58}\mbox{Ni}(^{36}\mbox{Ar},\alpha
\mbox{pn})^{88}\mbox{Tc}$ reaction at a beam energy of
$145~\mbox{MeV}$ was conducted at the VICKSI accelerator at the
HMI/Berlin. After an annihilation-$\gamma$-coincidence measurement
of the $7^-$ ($3350~\mbox{keV}$) and the $8^+$ ($3212~\mbox{keV}$)
yrast states in $^{88}\mbox{Mo}$, the ground state was suggested
to have a spin/parity of either $7^-$ or $8^+$ \cite{Rudo1991}. 
Gamma radiation in $^{88}\mbox{Mo}$ following the $\beta$ decay of a 
low-spin state in $^{88}\mbox{Tc}$ has not been observed in Ref. \cite{Weis1992}.
In a $\beta$-decay study at the Tandem Accelerator Laboratories at
Kyushu and Tsukuba using the
$^{58}\mbox{Ni}(^{32}\mbox{S},\mbox{pn})^{88}\mbox{Tc}$ reaction
\cite{Odah1996} two $\beta$-decaying states with half-lives
of $5.8(0.2)~\mbox{s}$ for a low-spin state $(2,3)^+$ and
$6.4(0.8)~\mbox{s}$ for a high-spin state ($6,7,8$) were observed
in coincidence with $\gamma$ rays. The $\beta$-decay endpoint
energies were measured by $\beta$-$\gamma$ coincidence gated on
the 741-keV $\gamma$ ray and the derived $Q_{\text{EC}}$
values of $^{88}\mbox{Tc}$ were given in two different papers
\cite{Odah1996,Shib1996}. In both cases these values are
significantly lower by $1.4~\mbox{MeV}$ and $2.2~\mbox{MeV}$, respectively, than those from the
systematic trends. Therefore, this $Q$ value has been replaced by
a systematic value in AME2003. Our mass excess value of
$-61679.1(3.8)~\mbox{keV}$ determined with JYFLTRAP is even less bound than the systematic value by $1030~\mbox{keV}$.
In analogy to the neighboring odd-odd nuclide $^{90}\mbox{Tc}$ we have most probably determined the mass 
of the high-spin state. Since the positions of
the ground and isomeric states are identical according to
extrapolations (see Table \ref{tab:NUBASE}) a treatment for an
unknown mixture of states \cite{Waps2003} cannot be applied and our result is
assigned as the ground state mass with an increased uncertainty 
of $87~\mbox{keV}$.\\

{\bf $^{90}\mbox{Ru}$}, {\bf $^{92}\mbox{Ru}$}, 
{\bf $^{92}\mbox{Rh}$}, and {\bf $^{93}\mbox{Rh}$}: The measurements reported here 
are the first experimental mass
determinations of these nuclides. The even-even ruthenium isotopes
have $0^+$ ground states and in $^{93}\mbox{Rh}$ no isomeric state was observed. All data identify the nuclei less bound
than suggested, based on the extrapolation from systematic trends
in AME2003 \cite{Audi2003a}. $^{92}\mbox{Rh}$ and $^{92}\mbox{Ru}$
are less bound by $361~\mbox{keV}$ and $111~\mbox{keV}$,
respectively. For $^{90}\mbox{Ru}$ a mass difference of
$427~\mbox{keV}$ deviates by 1.4$\sigma$ from the systematic
estimates of the AME2003. These observations
continue the trends in the $A = 90$ and $A = 92$ chains along the
isobaric distance from stability. $^{93}\mbox{Rh}$ is less bound
by $159~\mbox{keV}$. Here, the general observation of less bound
experimental mass values is continued.\\
Evidence for a possible existence of another long-lived state in
$^{92}\mbox{Rh}$ is given in a publication by Dean {\it et
al.} \cite{Dean2004}. Since observed feedings to the low-spin
($0^+$, $2^+$) levels in the daughter $^{92}\mbox{Ru}$ are larger
than expected when originating from the presumed ($\ge 6^+$) ground
state, a low-spin isomeric state might exist. Indeed, a second
half-life component was observed in the 866-keV $\gamma$-ray line, with a
deduced half-life of $0.53~\mbox{s}$. According to shell-model
calculations a $2^+$ state is predicted either about $50~\mbox{keV}$ 
\cite{Kast1997,Dean2004} or $211~\mbox{keV}$ \cite{Hern1997} below the $6^+$ state.\\
A total duration of our Penning trap measurement of $1.7~\mbox{s}$ resulted in 
a disintegration of most of the shorter-lived isomeric component, if 
it were produced, and our mass value is hence assigned to the high-spin isomer. In addition, 
the high-spin states are more favorably produced via heavy-ion fusion-evaporation reactions. 
In order to account for the unknown level scheme between the ($\ge 6^+$) and the 
$2^+$ states, the uncertainty of the ground-state mass excess value of $^{92}\mbox{Rh}$ is increased to 
$15~\mbox{keV}$. Spectroscopic experiments are required to confirm the low-energy level 
scheme of this nuclide which is of vital importance for the study of the two-proton decay 
of $^{94}\mbox{Ag}$ $(21^+)$ \cite{Mukh2005,Mukh2006,Roec2006,Pech2007,Kank2008}.\\

{\bf $^{94}\mbox{Ru}$}, {\bf $^{96}\mbox{Pd}$}: In both nuclides
the $0^+$ ground state and an $8^+$ isomeric state with an
excitation energy larger than $2.5~\mbox{MeV}$ were observed in previous 
measurements. Due to their $\mu$s half-lives, nuclides in the latter states have 
no influence on a cyclotron frequency measurement.\\
In AME2003 the ground state mass values are ultimately related to
the mass of the primary nuclide $^{96}\mbox{Ru}$. The mass of
$^{94}\mbox{Ru}$ was previously determined with an uncertainty of
$13~\mbox{keV}$ in the $^{96}\mbox{Ru}$(p,t) reaction
\cite{Ball1971}. The combined result from the Penning trap
experiments agrees within one standard deviation 
to the previously measured value.\\
The mass of $^{96}\mbox{Pd}$ was previously determined via its
$\beta$ decay \cite{Ryka1985} and the
$^{96}\mbox{Ru}(\mbox{p,n})^{96}\mbox{Rh}$ reaction
\cite{Ashk1970} $Q$ values. Its uncertainty of $150~\mbox{keV}$ is dominated 
by the decay measurement. According to our new experimental mass
value, with a 30-fold improved precision, the nucleus is by $51~\mbox{keV}$ less bound.\\

%%%%%%%%%%%%%%%%%%%%%%%%%%%%%%%%%%%%%%%%%%%%%%%%%%%%%%%%%%%%%%%%%%%%%%%%%%%%%%%%%%%%%%%%%%%%%%%%%%
%%%%%%%%%%%%%%%%%%%%%%%%%%%%%%%%%%%%%%%%%%%%%%%%%%%%%%%%%%%%%%%%%%%%%%%%%%%%%%%%%%%%%%%%ODD9/2%1/2
%%%%%%%%%%%%%%%%%%%%%%%%%%%%%%%%%%%%%%%%%%%%%%%%%%%%%%%%%%%%%%%%%%%%%%%%%%%%%%%%%%%%%%%%%%%%%%%%%%
Nuclides with assigned spin values {\bf $I^{\pi} = 9/2^+$} and {\bf $I^{\pi} = 1/2^-$} for the ground and 
first isomeric state, respectively: {\bf
$^{87}\mbox{Zr}$}, {\bf $^{89}\mbox{Mo}$}, {\bf $^{89}\mbox{Tc}$},
{\bf $^{91}\mbox{Tc}$}, {\bf $^{91}\mbox{Ru}$},
{\bf $^{93}\mbox{Ru}$}, and {\bf $^{95}\mbox{Rh}$}: In most cases the half-lives of the ground and
isomeric states are very similar or at least sufficiently long to allow for
a measurement of both states. The sole exception is
$^{89}\mbox{Mo}^m$, with a half-life of the excited isomer of
$190~\mbox{ms}$. The excitation energies of $^{89}\mbox{Tc}^m$ and
$^{91}\mbox{Ru}^m$ are well below the employed experimental
resolution of the present experiments of $\Delta m = 120~\mbox{keV}$ 
for SHIPTRAP or $\Delta m = 80 - 90~\mbox{keV}$ for
JYFLTRAP. Despite the fact that there is a difference of only four spin 
units between the isomeric state and the ground state, instead of seven for
$^{90}\mbox{Tc}$, we assume that the high-spin-state products are
favored due to the fusion-evaporation mechanism. This fact, 
combined with the results obtained in the count-rate class 
analysis for {$^{89}\mbox{Mo}$}, {$^{91}\mbox{Tc}$}, or
{$^{93}\mbox{Ru}$} which give no indication of an additional 
admixture by a different mass, supports the assignment of the mass
values of these species to the high-spin ground states.\\

{\bf $^{87}\mbox{Zr}$}: The result from JYFLTRAP agrees with the AME2003 value for the ground state,
which is derived from the mass of $^{90}\mbox{Zr}$ and the
$^{90}\mbox{Zr}(^{3}\mbox{He},^{6}\mbox{He})^{87}\mbox{Zr}$
reaction $Q$ value, determined at MSU \cite{Pard1978}. However,
our value deviates by 1.5 standard deviations from the result $ME =
-79384(28)~\mbox{keV}$ of an ESR Schottky measurement
\cite{Litv2005}, which is not used in the adjustment procedure of
the Atomic Mass Evaluation due to its higher uncertainty.\\

{\bf $^{89}\mbox{Mo}$}: The value from JYFLTRAP agrees with the
AME2003 value, which is determined via a reaction $Q$-value measurement 
$^{92}\mbox{Mo}(^{3}\mbox{He},^{6}\mbox{He})^{89}\mbox{Mo}$ from a
study at MSU \cite{Pard1980} and the mass of the primary nuclide
$^{92}\mbox{Mo}$.\\

{\bf $^{89}\mbox{Tc}$}: For $^{89}\mbox{Tc}$ a $Q_\beta$ 
measurement was performed by Heiguchi {\it et al.} \cite{Heig1991}
yielding a mass excess of $ME = -67500(210)~\mbox{keV}$. However,
AME2003 assumed this value to be lower by $350~\mbox{keV}$ than given 
by this experiment due to its difference to other experimental results in this 
region. Our combined result of $ME =
-67394.8(3.7)~\mbox{keV}$ lies $445~\mbox{keV}$ above the 
value suggested by systematics in AME2003 and is in full agreement with
\cite{Heig1991}, although 30 times more precise. Since a pure production 
of the high-spin state is assumed, a correction for an isomeric 
mixture, $E_{\text{ex}} = 62.6~\mbox{keV}$ \cite{Rudo1995}, is 
not applied.\\

{\bf $^{91}\mbox{Tc}$}: This nuclide has an isomeric state with a
well-known excitation energy of $E_{\text{ex}} = 139.3(3)~\mbox{keV}$, which is sufficiently large to 
be resolved in the present measurements at SHIPTRAP and JYFLTRAP. 
Since no count-rate-dependent effects were observed, we assume the
pure presence of ions in the $9/2^+$ ground state. The $Q_\beta$-decay 
energy of $^{91}\mbox{Tc}$ produced in a (p,2n) reaction at
$30~\mbox{MeV}$ beam energy \cite{Iafi1974} resulted in a mass excess
value of $-75980(200)~\mbox{keV}$, which is in good agreement with
our result, but 60 times less precise.\\

{\bf $^{91}\mbox{Ru}$}: The mass of $^{91}\mbox{Ru}$ was
previously unknown. In AME2003 it was deduced from the known mass
of its isomer, using an estimated excitation energy of
$80(300)\#~\mbox{keV}$. Due to the high uncertainty in the
isomeric excitation energy, it cannot be definitely concluded 
that ground and isomeric states were resolved in the measurement,
although a mass resolution of $\Delta m = 80~\mbox{keV}$ was
employed at JYFLTRAP. Here, no count-rate-dependent effects were 
observed in the analysis. 
Our averaged result for $^{91}\mbox{Ru}$ combined with the mass of
the isomer, which was obtained by Hagberg \cite{Hagb1983} from
delayed proton-($\epsilon p$)-decay-energy measurements, now locates 
the excited state at -340(500) keV. Given the large
uncertainty in this value the assignment of the isomeric state in
this nuclide is still ambiguous. However, Hagberg clearly states
that the value they obtained for the ($\epsilon p$) decay energy
was only a lower limit. If we assume this energy to be by
$500~\mbox{keV}$ higher than the actual value of
$4300(500)~\mbox{keV}$ \cite{Hagb1983}, then the excitation energy would be $+160~\mbox{keV}$.\\

{\bf $^{93}\mbox{Ru}$}: In this nuclide the two lowest states have
sufficiently long half-lives for both to be delivered to the JYFLTRAP
Penning trap setup. Since the excitation energy is $734.4(1)~\mbox{keV}$ and 
precisely known, the ground and isomeric states can 
in principle be resolved 
with a mass resolution $\Delta m$ of $85~\mbox{keV}$.\\
The derived mass excess value agrees well with the value for the
$9/2^+$ ground state, which is determined from a $Q_{\beta}$
measurement via $^{93}\mbox{Tc}$ and a (p,$\gamma$)-reaction $Q$
value to the stable nuclide $^{92}\mbox{Mo}$ \cite{Äyst1983}. The
uncertainty in this mass value is now improved by more than a factor of 20.\\

{\bf $^{95}\mbox{Rh}$}: This nuclide has been studied in
$\gamma$-$\gamma$ and $\beta$-$\gamma$ coincidence measurements at the
McGill synchrocyclotron \cite{Weif1975}. A 5-min ground state and
a 2-min isomeric state have been observed. Additionally, the isomeric
transition between the $1/2^-$ and the $9/2^+$ states has been
identified. From the $\beta$-endpoint energy, the $Q$ value and the mass
excess have been derived with an uncertainty of $150~\mbox{keV}$.
In the $^{40}\mbox{Ca}$ + $^{58}\mbox{Ni}$ reaction, as employed
in this experiment, the high-spin state is more likely to
be produced. This assumption is supported by the study of neutron-deficient ruthenium and
rhodium isotopes produced with the same reaction at
the Munich tandem and heavy-ion post accelerator \cite{Nolt1980a}.
In that work, the $1/2^-$ state has been only weakly populated at a beam
energy of $135~\mbox{MeV}$, whereas a strong production of the
$9/2^+$ ground state has been observed.\\
Similar to the situation of $^{93}\mbox{Ru}$, the mass value
determined by JYFLTRAP is in excellent agreement with the previous
value of the $9/2^+$ ground state, but now with more than
35-fold lower uncertainty compared with the previous result
stemming from $Q_{\beta}$ decay to the primary nuclide
$^{95}\mbox{Ru}$ \cite{Pins1968}.\\

{\bf $^{92}\mbox{Tc}$}: Our averaged mass excess value of $ME =
-78924.7(3.7)~\mbox{keV}$ is found to be in good agreement with
all former measurements from studies using either (p,n) or
($^{3}\mbox{He}$,t) reactions \cite{Moor1966,Haya1973}.\\

\begin{figure}
\includegraphics[width=0.48\textwidth]{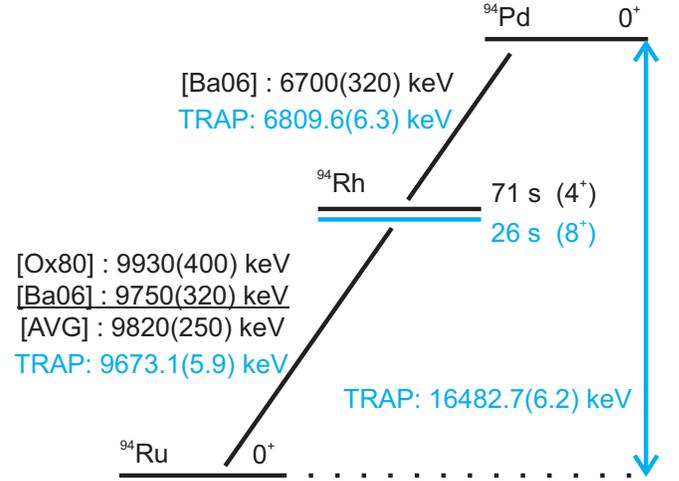}
\caption{\label{94Rh}$Q_{\text{EC}}$ values of the $^{94}\mbox{Pd}$ and $^{94}\mbox{Rh}$ decays, measured at the Foster Radiation
Laboratory (Ox80) \cite{Oxor1980}, the GSI on-line mass separator (Ba06) \cite{Bati2006}, and within this work (TRAP). Both decay experiments studied the 71-s 
($4^+$) isomeric component and their weighted average is given. The $Q_{\text{EC}}$ values from JYFLTRAP for these $A = 94$ nuclides involve the $(8^+)$ isomeric state in $^{94}\mbox{Rh}$.}   
\end{figure}

%%%%%%%%%%%%%%%%%%%%%%%%%%%%%%%%%%%%%%%%%%%%%%%%%%%%%%%%%%%%%%%%%%%%%%%%%%%%%%%%%%%%%%%%%%%%%%%%%%
%%%%%%%%%%%%%%%%%%%%%%%%%%%%%%%%%%%%%%%%%%%%%%%%%%%%%%%%%%%%%%%%%%%%%%%%%%%%%%%%%%%%%%%%%%%RHODIUM
%%%%%%%%%%%%%%%%%%%%%%%%%%%%%%%%%%%%%%%%%%%%%%%%%%%%%%%%%%%%%%%%%%%%%%%%%%%%%%%%%%%%%%%%%%%%%%%%%%
{\bf $^{94}\mbox{Rh}$}: In this nuclide, two states are reported in NUBASE (see~Table~\ref{tab:NUBASE}) with 
tentatively assigned spin values of $(2^+, 4^+)$ and $(8^+)$ for the 71-s ground and 26-s isomeric states, respectively. 
Their relative order is, however, uncertain since the ground-state mass value is only
estimated from systematic trends and the one of the excited isomeric state is determined via the $Q$ value to
$^{94}\mbox{Ru}$ \cite{Waps2003}, based on a $\beta$-endpoint measurement for the 71-s low-spin state, referred to $(3^+)$ in Ref. \cite{Oxor1980}. 
Despite the high uncertainty in the excitation energy of $E = 300(200)\#~\mbox{keV}$, a value which is estimated from systematic trends \footnote{Note that a possible lower value in the excitation energy of $^{90}\mbox{Tc}^m$ will affect this estimate.}, we can assume to have resolved both states with a mass resolution of $\Delta m = 86~\mbox{keV}$.\\
An experiment studying $\beta$-delayed proton emission, conducted at the on-line mass separator \cite{Roec2003} at
GSI/Darmstadt, observed that the 26-s $^{94}\mbox{Rh}$ activity did not show any 
grow-in effect \cite{Kurc1982}. It was deduced that this state was probably directly produced in the 
employed $^{40}\mbox{Ca}$ + $^{58}\mbox{Ni}$ reactions and released from the ion source, whereas
the observed low-spin 71-s state was fed from the $\beta$ decay of $^{94}\mbox{Pd}$. In addition, a recent study on low-lying 
levels in $^{94}\mbox{Ru}$ following the identical reaction \cite{Mill2007} concluded that the $(4^+)$ state had not been significantly 
populated. Hence, we tentatively assign our measured mass value to the $(8^+)$ high-spin state.\\
A recent $Q_{\text{EC}}$-value determination of $^{94}\mbox{Pd}$ and of the 71-s state in $^{94}\mbox{Rh}$ was carried out at 
the GSI on-line mass separator  after fusion-evaporation reactions ($^{40}\mbox{Ca}$ on $^{58}\mbox{Ni}$) employing the 
Total Absorption Spectrometer (TAS) \cite{Bati2006}. Here, the activity of the latter state was produced as a daughter of the
$^{94}\mbox{Pd}$ $\beta$ decay.\\
The results on $Q_{\text{EC}}$ values of the $^{94}\mbox{Pd}$ and $^{94}\mbox{Rh}$ decays are 
summarized in Fig. \ref{94Rh}. Both decay experiments studied the 71-s component and their weighted average is 
given. The summed $Q_{\text{EC}}$ values for both nuclides, derived from JYFLTRAP mass values and involving the high-spin $(8^+)$ state in 
$^{94}\mbox{Rh}$, agree with the mass difference between $^{94}\mbox{Pd}$ and $^{94}\mbox{Ru}$. The comparison of these data indicate an energetic ordering with an $(8^+)$ ground state, in accordance with shell-model calculations \cite{Hern1997}. Although this comparison is governed by large uncertainties in the $Q$-value determinations, we 
tentatively assign our result to the ground state. A similar conclusion was already drawn in Ref. \cite{Bati2006} after a comparison 
with the mass excess value by the Canadian Penning Trap for $^{94}\mbox{Rh}$, 
using fusion-evaporation reactions with heavy-ion beams at ANL \cite{Clar2005}.\\ 

%%
%%%%%%%%%%%%%%%%%%%%%%%%%%%%%%%%%%%%%%%%%%%%%%%%%%%%%%%%%%%%%%%%%%%%%%%%%%%%%%%%%%%%%%%%%%%%%%%%%%
%%%%%%%%%%%%%%%%%%%%%%%%%%%%%%%%%%%%%%%%%%%%%%%%%%%%%%%%%%%%%%%%%%%%%%%%%%%%%%%%%%%%%%%%%PALLADIUM
%%%%%%%%%%%%%%%%%%%%%%%%%%%%%%%%%%%%%%%%%%%%%%%%%%%%%%%%%%%%%%%%%%%%%%%%%%%%%%%%%%%%%%%%%%%%%%%%%%
{\bf $^{94}\mbox{Pd}$}: Since the excited isomeric state possesses
a half-life in the nanoseconds range, a clean sample of ions in
the ground state has been delivered to the experiment. The mass of
this $0^+$ state with a half-life of $9~\mbox{s}$ is determined
experimentally for the first time. It agrees with the value of the
AME2003 derived from the extrapolation of
systematic experimental trends.\\

{\bf $^{95}\mbox{Pd}$}: The nuclide $^{95}\mbox{Pd}$ has been
produced in the fusion-evaporation reaction
$^{\text{nat}}\mbox{Ni}(^{40}\mbox{Ca},2\mbox{pxn})^{95}\mbox{Pd}$ with a
predicted cross section \cite{PACE2005} on the order of a mb
at an energy about $145~\mbox{MeV}$. Since the RF-excitation time for the radial motions in
the trap was $800~\mbox{ms}$ and the half-lives are sufficiently long, nuclides in their ground 
and isomeric states were observed in the measurement. Indeed, cyclotron resonance curves at two different
frequencies were recorded.\\
The long-lived isomeric state $^{95}\mbox{Pd}^{m}$ with a half-life
of $13.3(3)~\mbox{s}$ was observed in investigations of
$\beta$-delayed proton emission \cite{Nolt1980b} and $\beta$ decay
\cite{Kurc1982}. The spin value of this state whose $\beta$-p
chain strongly feeds the $8^{+}$ state in $^{94}\mbox{Ru}$ was
assigned as $21/2^{+}$ \cite{Nolt1980b} in accordance with the
shell-model predictions \cite{Ogaw1983}. The interpretation of
this state as a spin-gap isomer is based on the long-lived
character of the decay \cite{Nolt1980b} and on the observation of
the weak palladium K X-rays at mass $A = 95$ \cite{Kurc1982}.\\
%Internal decay from $^{95}\mbox{Pd}^{m}$ giving the
%exact energy position of this state was not yet observed, and hence the spin-gap 
%character of this $21/2^{+}$ state was not confirmed. 
Moreover, the
exact position of the $21/2^+$ state will determine the energies of
high-spin yrast states up to $43/2^+$ built on this isomer
\cite{Doer2003,Arne1994}. The value for the ground-state mass
excess is $-69961.6(4.8)\mbox{keV}$ and for the isomeric state it
is $-68086.2(4.7)~\mbox{keV}$. Using these directly measured
values we determined the excitation energy of the isomeric state as $E
= 1875.4(6.7)~\mbox{keV}$. This is the first direct mass determination of 
such a high spin state. The value is in excellent agreement
with the one of $1875~\mbox{keV}$, obtained from $\gamma$-decay
spectroscopy \cite{Doer2003}. Our new experimental result can be compared with 
the theoretical predictions of $1.9~\mbox{MeV}$
\cite{Ogaw1983}, $1.8~\mbox{MeV}$ \cite{Arne1994},
$1.97~\mbox{MeV}$ \cite{Schm1997} and with evaluated data
$1.86(50)\#~\mbox{MeV}$ of the AME2003 \cite{NUBA2003}. Recently,
the energy of this isomeric state was determined from the
$\gamma$ transitions which connect the states built on the ground
and isomeric states \cite{Marg2004} to $1876~\mbox{keV}$, in agreement with our value. The latter 
work concludes that the $21/2^{+}$ state lies lower than the $15/2^{+}$, $17/2^{+}$, and
$19/2^{+}$ states, manifesting its spin-gap character of isomerism. Hence, the excitation 
energy confirms the spin-gap character expected from the shell-model calculations taking into 
account the inert $^{100}\mbox{Sn}$ core and, therefore, indirectly confirming the 
magic character of the nuclide with $Z = N = 50$.\\
The $Q_{\text {EC}}$ value for the isomeric state is equal to
$10256.1(6.3)~\mbox{keV}$. The $\beta$ decay of the isomeric state
dominantly feeds the 2449-keV~($21/2^+$) level in $^{95}\mbox{Rh}$
\cite{Kurc1982}, which results in a transition $Q$ value of $8829(6.3)~\mbox{keV}$. Taking into account a decay 
branching of 35.8\% \cite{Kurc1982}, we obtain a log {\it {ft}} = 5.5, 
confirming the allowed character of the transition.\\
%%%%%%%%%%%%%%%%%%%%%%%%%%%%%%%%%%%%%%%%%%%%%%%%%%%%%%%%%%%%%%%%%%%%%%%%%%%%%%%%%%%%%%%%%%%%%%%%%%
%%%%%%%%%%%%%%%%%%%%%%%%%%%%%%%%%%%%%%%%%%%%%%%%%%%%%%%%%%%%%%%%%%%%%%%%%%%%%%%%%NUCLEAR STRUCTURE
%%%%%%%%%%%%%%%%%%%%%%%%%%%%%%%%%%%%%%%%%%%%%%%%%%%%%%%%%%%%%%%%%%%%%%%%%%%%%%%%%%%%%%%%%%%%%%%%%%
\section{\label{nuc}Observations in Nuclear Structure}
\begin{figure*}
\includegraphics[width=0.70\textwidth]{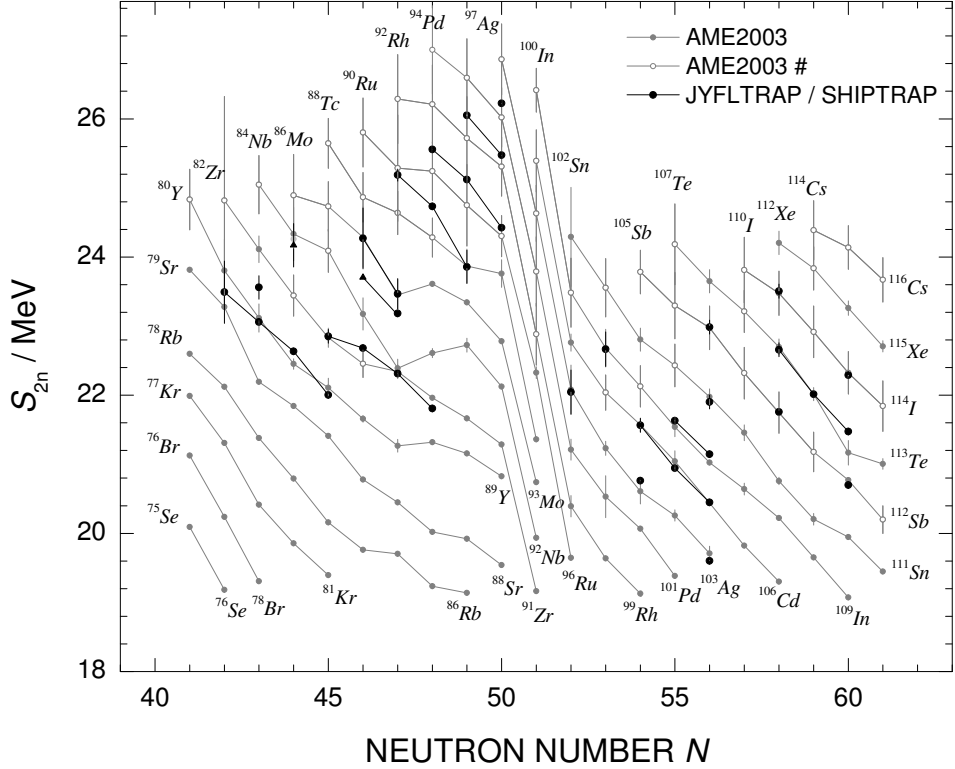}
\caption{\label{S2n}Two-neutron separation energies $S_{\text{2n}}$ derived from this work (black circles) and AME2003 (grey circles). 
The data points of AME2003 are either based solely on experimental results (filled circles, AME2003) or at least partly 
derived from extrapolations of systematic trends (open circles, AME2003 \#). The data points for $^{85}\mbox{Nb}$ and $^{87}\mbox{Nb}$ (triangles) imply a correction due to the presence of an isomeric state. For a detailed discussion see text.}
\end{figure*}

\begin{figure*}
\includegraphics[width=0.70\textwidth]{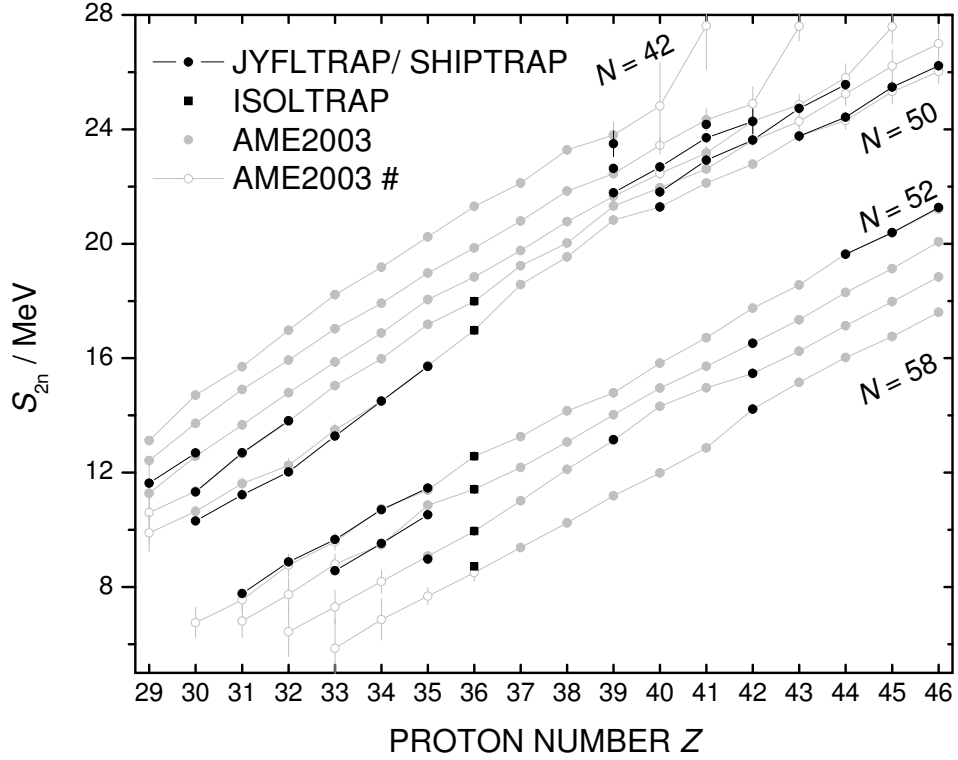}
\caption{\label{shell gap}Two-neutron separation energies $S_{\text{2n}}$ as a function of the proton number for even 
isotones across the $N = 50$ shell gap. The data points of AME2003 are either based solely on experimental results (filled circles, AME2003) or at 
least partly derived from extrapolations of systematic trends (open circles, AME2003 \#). The data of krypton isotopes ($Z = 36$) are from Ref. \cite{Dela2006} (black squares). The data points for $^{85}\mbox{Nb}$ and $^{87}\mbox{Nb}$ imply a correction due to the possible presence of an isomeric state (compare Fig. \ref{S2n}).}
\end{figure*}

\begin{figure*}
\includegraphics[width=0.90\textwidth]{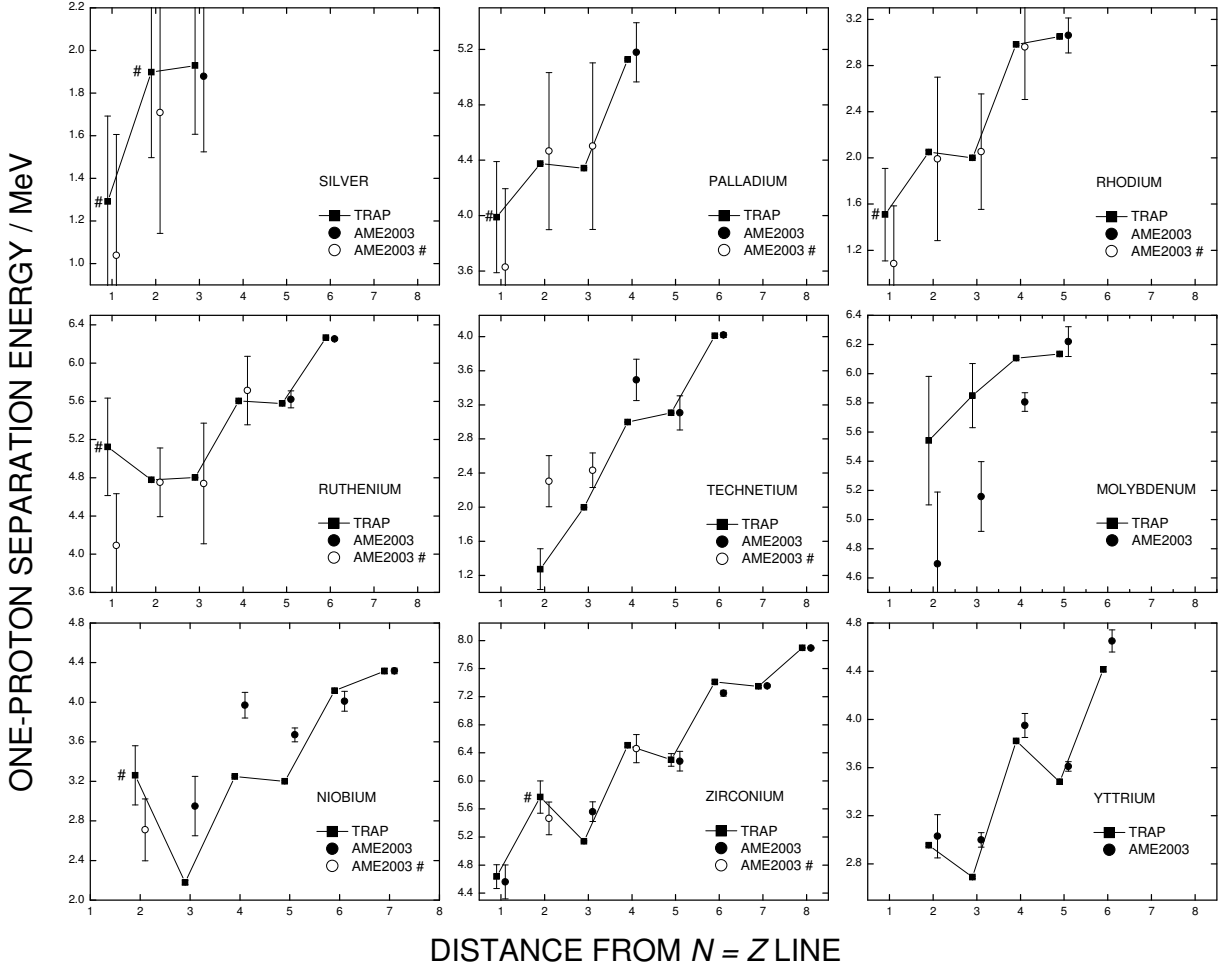}
\caption{\label{Sp_iso}One-proton separation energies as a function of the distance from the $N = Z$ line for the 
isobars from yttrium to silver. The data points of AME2003 are either based solely on experimental results (filled 
circles, AME2003) or at least partly derived from extrapolations of systematic trends (open circles, AME2003 \#). The trap data result 
in major adjustments of some of the values and moreover in a considerable reduction of the error bars. Separation energies $S_p$ comprising 
data from $^{83}\mbox{Y}$ or $^{85,87,88}\mbox{Nb}$ were corrected for the possible presence of an unknown isomeric contribution. For some of 
the nuclides the separation energies are a combination of a trap measurement and a value of the AME2003 and hence exhibit a larger uncertainty. 
These are $^{80-84}\mbox{Y}$, $^{81,82,86-88}\mbox{Zr}$, $^{84,89}\mbox{Nb}$, $^{86,87}\mbox{Mo}$, $^{88,91,92}\mbox{Tc}$, 
$^{89,94}\mbox{Ru}$, $^{91}\mbox{Rh}$, $^{93}\mbox{Pd}$, and $^{95-97}\mbox{Ag}$. In addition, data combined with extrapolated AME2003 values are marked 
by `\#'.}
\end{figure*}

The two-neutron separation energy $S_{\text{2n}} = B(Z,N) - B(Z,N-2)$ is the mass 
derivative which best represents the systematic behaviour of the studied binding energies and the nuclear structure evolution 
around the well-established $N = 50$ shell closure. Figure \ref{S2n} shows the $S_{\text{2n}}$ values of the isotopes from selenium to cesium. The data 
available in AME2003 (grey circles) have been complemented by new experimental data (black circles) from JYFLTRAP \cite{Kank2006}, 
SHIPTRAP \cite{Mart2007} and this work. For the lowest neutron numbers, data points are partly combined with AME2003 values 
that are based on experimental input data. In addition to the expected decrease of the $S_{\text{2n}}$ energies with increasing 
neutron number the well-established $N = 50$ shell closure is observed.\\
The region with neutron numbers $N \ge 50$, which was previously partly established by experimental results, is well reproduced by the 
data from Mart\'in {\it et al.} \cite{Mart2007}. In contrast, the region below this neutron-shell closure, which was based mostly on 
extrapolations of systematic trends, is substantially modified for isotopes from technetium to palladium. The separation energies are now solely defined by experiment and all values are shifted consistently towards higher energies 
comprising equidistant lines of similar curvatures.\\ 
Comparing the changes in the two-neutron separation energies for the isotopes of niobium and molybdenum at neutron numbers $N = 46$ and $N = 47$ \footnote{A correction due to an unknown contribution of isomeric states has been applied for the data from Ref. \cite{Kank2006}.} two observations are made: (1) both experimental data points of the AME2003 for molybdenum are still well reproduced after including our mass values for $^{88}\mbox{Mo}$ and $^{89}\mbox{Mo}$, (2) a shift of the separation energies towards higher values is observed in niobium. The mass values of these nuclides are now entirely determined by JYFLTRAP data, whereas the previous AME2003 data were based on $Q_{\beta}$ measurements. Note that the contribution of known, but unobserved, isomers has been accounted.\\
The $S_{\text{2n}}$ data point for $^{85}\mbox{Nb}$ at $N = 44$ is calculated using the masses of the nuclides $^{85}\mbox{Nb}$ and $^{83}\mbox{Nb}$, which were previously mainly determined by sequences of $\beta$-decay endpoint measurements up to degrees \footnote{Number of successive links indicating the distance to the primary data set in the atomic mass evaluation.} 5 and 4, respectively 
\cite{NUBA2003}. The possibility of too low mass values due to unobserved $\gamma$ lines can in theory be attributed to any of 
the contributing results. Triggered by an inconsistency in the $Q_{\beta}$ value for $^{85}\mbox{Nb}$ given in Ref.
\cite{Kuro1988}, the AME2003 mass value was recalculated and is now shifted by 70 keV to a less bound value (light grey circle).\\
Furthermore, the presence of long-lived isomeric states can obscure the interpretation of results from mass 
measurements. Along either of the $A = 83$ and $A = 85$ chains two isobars, $^{83}\mbox{Zr}$, $^{83}\mbox{Y}$ and $^{85}\mbox{Nb}$, $^{85}\mbox{Zr}$, 
were studied with the JYFLTRAP mass spectrometer, but due to the open question of the isomeric mass-to-state assignment for 
$^{83}\mbox{Y}$ and $^{85}\mbox{Nb}$  \cite{Kank2006} a correction for a mixture between two states has been applied.\\ 
Departing from stability along the $A = 83$ chain, the mass values from JYFLTRAP identfy the nuclei less bound by 130 and 
552 keV for $^{83}\mbox{Y}$ and 
$^{83}\mbox{Zr}$, respectively. 
Hence, the mass value for $^{83}\mbox{Nb}$ is also expected to be less bound by at least 
$682~\mbox{keV}$ compared with the present AME2003 value. For the isotope containing two more neutrons, $^{85}\mbox{Nb}$, the question of an isomeric mass-to-state assignment is not solved and a state with an isomeric excitation energy of $69~\mbox{keV}$ is reported in Ref. \cite{Kank2006}. These are considered and combined with a modified mass value for $^{83}\mbox{Nb}$ and illustrated by a black upfacing triangle, 
in agreement with 
the rather large uncertainty of the AME2003 value determined by Ref. \cite{Kuro1988}. Due to these open questions of isomerism this region should be addressed in a future measurement for final unambiguous mass-to-state assignments. Ideally, the cyclotron frequency determinations should be assisted by either a hyperfine-state-selective laser ionization \cite{VanR2004,Blau2004} or a nuclear decay-spectroscopy experiment \cite{Rint2007}.\\ 
In Fig. \ref{shell gap} the two-neutron separation energies $S_{\text{2n}}$ are displayed as a function of the proton number 
for even isotones across the $N = 50$ shell 
gap. This gap $\Delta_n (N = 50)$ is defined as the difference between the two-neutron separation energies for $N = 50$ and 
$N = 52$, noticeable in Fig. \ref{S2n} as the steepness of the slope and in Fig. \ref{shell gap} as a gap between $N = 50$ and $N = 52$. In 
addition to the neutron-deficient data described in this work, Penning trap data from JYFLTRAP \cite{Haka2008,Raha2007,Raha2007a} and ISOLTRAP \cite{Dela2006} contribute to delineate the position of the shell gap on the neutron-rich side. A minimum in the shell-gap energy was observed for $Z = 32$ in Ref. \cite{Haka2008} with an indication of an enhancement towards the doubly-magic $^{78}\mbox{Ni}$ ($Z = 28$).\\
In the lower set of lines above the magic shell closure, a very regular behavior is observed for isotopes 
beyond molybdenum ($Z \ge 42$), indicating the rigidity of the $N = 50$ shell gap towards $^{100}\mbox{Sn}$, whereas an increasing difference develops between the $N = 56$ and $N = 58$ isotones below niobium ($Z \le 41$). This 
is in accordance with the developing shape change at $N = 59$ for niobium, zirconium \cite{Lher1994,Camp2002}, and yttrium \cite{Chea2007}. Contrary to this, the regular line spacing is visibly compressed for the yttrium isotopes at mid-proton shell ($Z = 39$).\\   
Many of the studied nuclides play a role in the astrophysical rp- and $\nu$p-processes and their proton-capture rates and 
finally the isotonic abundance ratios are influenced by our mass data (see Fig. \ref{chart_JYFL_SHIP}). In consequence of these measurements, 
which are partly the first experimental determinations, 
the one-proton separation energies are now more accurately known. This is illustrated in 
Fig. \ref{Sp_iso} for elements from yttrium to silver as a function of the isotopic distance from the $N = Z$ line. The one-proton separation energies $S_p$ of 
more than 45 nuclides have been modified by the new experimental data and those from Ref. \cite{Kank2006}. Since these often affect both mass values, $m(A,N)$ and $m(A-1,N)$, by a similar amount, the resulting corrections in the separation energies are partly canceled out. Nevertheless, the reduction of experimental uncertainties by more than a factor of ten provides a rectified set of input data for astrophysical calculations and will result in a more reliable modeling of the pathways, since in previous network calculations values with large uncertainties and assumptions based on extrapolations have been used. 
\clearpage
%%%%%%%%%%%%%%%%%%%%%%%%%%%%%%%%%%%%%%%%%%%%%%%%%%%%%%%%%%%%%%%%%%%%%%%%%%%%%%%%%%%%%%%%%%%%%%%%%%
%%%%%%%%%%%%%%%%%%%%%%%%%%%%%%%%%%%%%%%%%%%%%%%%%%%%%%%%%%%%%%%%%%%%%%%%%%%%%%%%%%%%%%%%%%%%%ASTRO
%%%%%%%%%%%%%%%%%%%%%%%%%%%%%%%%%%%%%%%%%%%%%%%%%%%%%%%%%%%%%%%%%%%%%%%%%%%%%%%%%%%%%%%%%%%%%%%%%%
\section{\label{ast}Implications for Astrophysics}
The neutron-deficient nuclei measured in this work are located in the paths 
of two astrophysical nucleosynthesis processes: the rp-process 
\cite{Wallace.Woosley:1981,Scha2001}
and the recently discovered $\nu$p-process 
\cite{Froh2006,Prue2006,Wana2006} 
(see Fig. \ref{chart_JYFL_SHIP}).
In the following, we will investigate the implications of the 
newly-determined mass values on the reaction flow and the nucleosynthesis
yields of the $\nu$p-process.
A similar investigation for the rp-process will be subject of a forthcoming
paper.

The $\nu$p-process occurs in explosive environments when proton-rich
matter is ejected under the influence of strong neutrino fluxes. This
includes the inner ejecta of core-collapse supernova
\cite{Buras.Rampp.ea:2006,Liebendoerfer.Mezzacappa.ea:2001a,Thompson.Quataert.Burrows:2005}
and possible ejecta from black hole accretion disks in the collapsar 
model of gamma-ray bursts \cite{Surman.McLaughlin.ea:2008}.
The matter in these ejecta is heated to temperatures well above 10~GK
and becomes fully dissociated into protons and neutrons.
The ratio of protons to neutrons is mainly determined by neutrino and 
antineutrino absorptions on neutrons and protons, respectively. As the matter 
expands and cools, the free neutrons and protons start to form 
$\alpha$ particles.
Later, at temperatures around 5~GK, $\alpha$ particles assemble into heavier 
nuclei but the expansion of matter is so fast that only a few iron-group 
nuclei are formed.
Once the temperature reaches 
around 2~GK the composition of the ejecta consists -- in order of decreasing 
abundance -- mostly of $^4$He, protons, and iron group nuclei with $N\sim Z$
(mainly $^{56}$Ni).

Without neutrinos, the synthesis of nuclei beyond the iron peak becomes very 
inefficient due to bottleneck nuclei (mainly even-even $N=Z$ nuclei) with long 
$\beta$-decay half-lives and small proton-capture cross sections.
However, during the expansion the matter is subject to a large neutrino and
antineutrino flux from the proto-neutron star.

While neutrons are bound in neutron-deficient $N = Z$ nuclei and neutrino 
captures on these nuclei are negligible due to energetics, antineutrinos are readily captured both on free protons and on heavy nuclei
on a timescale of a few seconds. As protons are more abundant than heavy nuclei
antineutrino captures occur predominantly on protons, leading to residual 
neutron densities of $10^{14}$--$10^{15}$~cm$^{-3}$ for several seconds.
These neutrons are then easily captured by the heavy neutron-deficient nuclei,
for example $^{64}$Ge, inducing (n,p) reactions with time scales much shorter 
than the $\beta$-decay half-life.
Now proton captures can occur on the produced nuclei, allowing the 
nucleosynthesis flow to continue to heavier nuclei.
The $\nu$p-process \cite{Froh2006} 
is this sequence of (p,$\gamma$) reactions followed by (n,p) or 
$\beta^+$-decays, where the neutrons are supplied by antineutrino captures 
on free protons.

Here, the impact of the new mass measurements on the $\nu$p-process 
nucleosynthesis is studied in post-processing calculations
of a representative trajectory from the explosion of a 15~M$_{\odot}$ star 
\cite{Janka.Buras.Rampp:2003}.  This trajectory is characterized by 
efficiently synthesizing nuclei with $A>90$ and is the same as used in Refs.\ 
\cite{Prue2006,Fisker.Hoffman.Pruet:2007}.
The neutrino temperatures and luminosities are assumed to remain constant 
at the values reached  at ~$1000$~ms  postbounce. With this assumption our 
results are directly comparable to those from Ref.\ 
\cite{Prue2006}.

Two sets of astrophysical reaction rates were used in the reaction network.
A reference set \cite{Rauscher:2008}
included theoretical rates obtained with a new version of
the NON-SMOKER code \cite{Rauscher.Thielemann:1998}. The theory is similar to 
previous
calculations \cite{Rauscher.Thielemann:2000} but containing a number of
updates and improvements, among which the latest information on excited states
\cite{nndc} and masses from the AME2003 \cite{Audi2003a} are the most important in the current
context. To consistently study the impact of the new mass data, a second rate
set was prepared, similar to the reference set except for the inclusion of
mass values from this work and Ref. \cite{Kank2006}. 
It has to be noted that we did
not only change the reaction $Q$ values but consistently recalculated the
theoretical rates with the newly implied separation energies.

\begin{figure}
\includegraphics[width=0.48\textwidth]{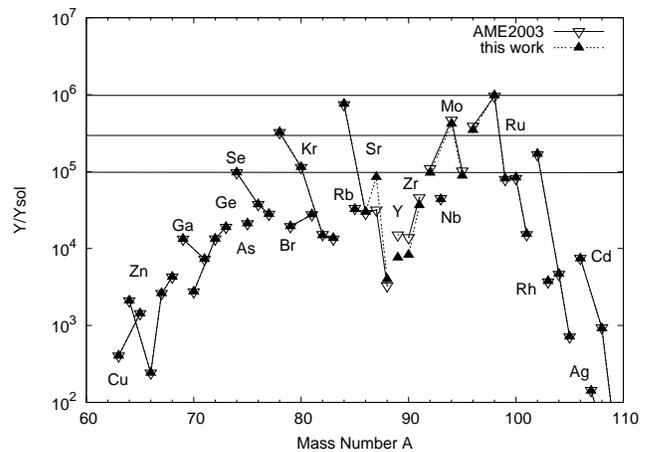}
\caption{\label{fig:ay}Final abundances after decay to stability relative to solar 
abundances \cite{Lodders:2003}. Filled upfacing triangles are for a calculation 
using the masses studied here. Open downfacing triangles are for a calculation 
using AME2003 masses.}
\end{figure}

Figure \ref{fig:ay} shows the final abundances normalized to solar abundances
after decay to stability for the two sets of thermonuclear reaction rates. 
Only nuclei produced in the p-rich ejecta are shown. As is clearly seen from the
figure there is no difference in the yields for the two different sets of 
rates except for a few nuclei in the mass range $85 < A < 95$, namely for
$^{87,88}$Sr, $^{89}$Y, and $^{90,91}$Zr. This can be directly traced back
to the large change in the mass of $^{88}$Tc 
($\Delta ME_{\mathrm{AME2003-TRAP}}=-1031$~keV, see Table~II).
This change in mass leads to an increase in the reaction rate for 
$^{88}$Tc$(\gamma,\mbox{p})^{87}$Mo at the relevant temperatures and therefore 
a relative suppression of the flow through $^{87}$Mo(p,$\gamma$)$^{88}$Tc.

\begin{figure}
\includegraphics[width=0.48\textwidth]{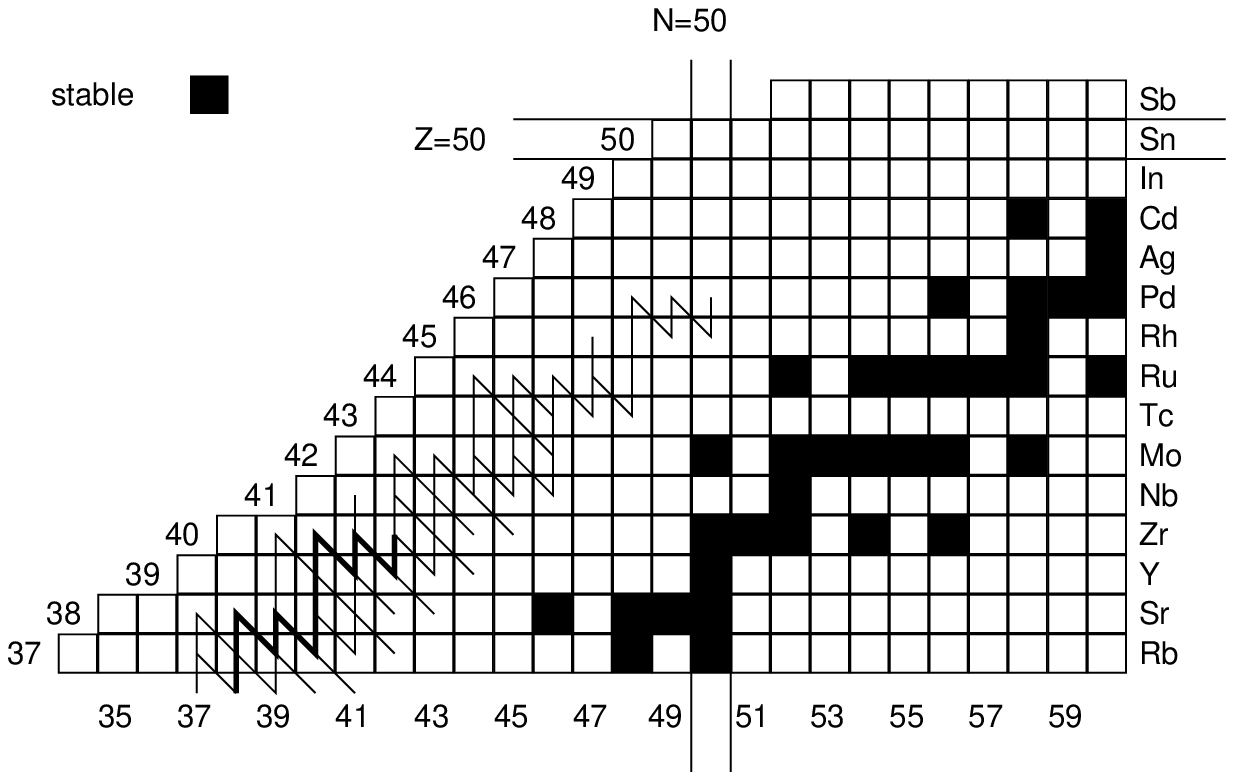}
\caption{\label{fig:flux-trp8} The time-integrated reaction flow for the 
$\nu$p-process using the reaction rate set based on the AME2003 mass data only.
The reaction flows shown are more than 10\% (thick line) and 1--10\% 
(thin line) of the reaction flow through the triple-$\alpha$-reaction.}
%\end{figure}
%
%
%\begin{figure}
\includegraphics[width=0.48\textwidth]{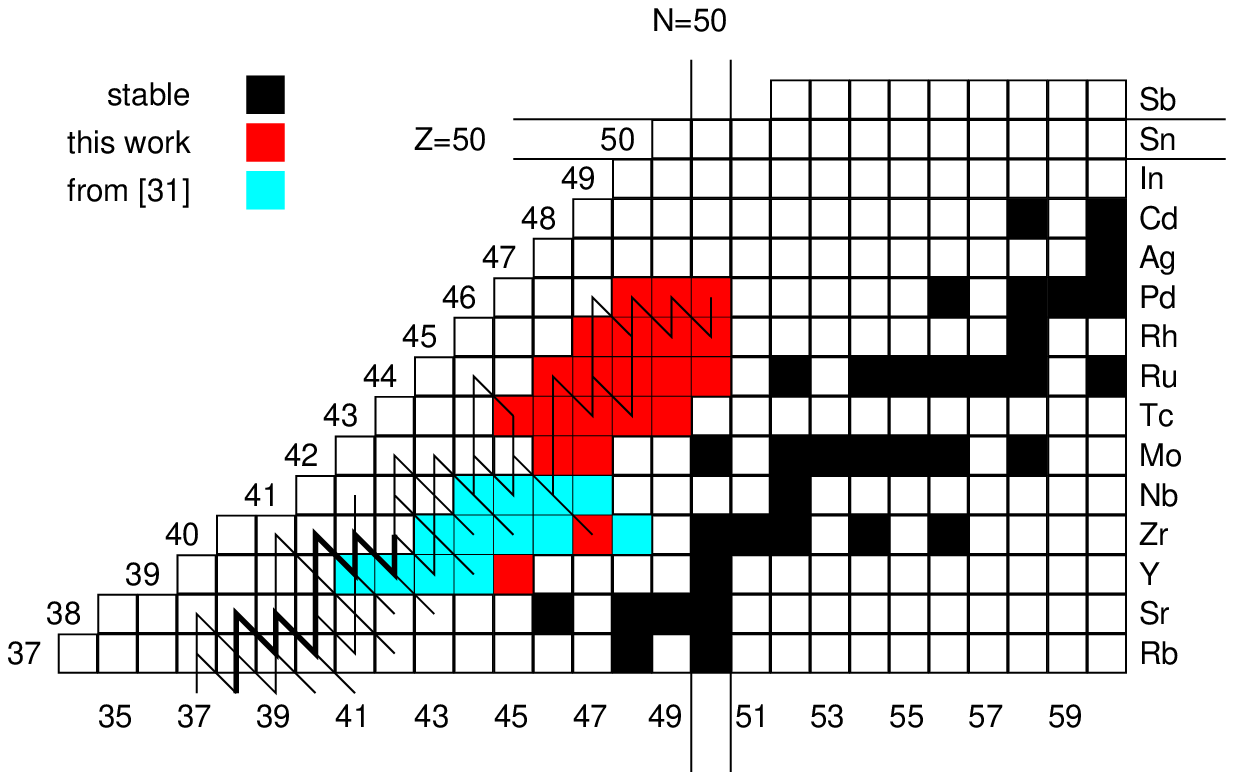}
\caption{\label{fig:flux-jytr} The time-integrated reaction flow for the 
$\nu$p-process using the reaction rate set with the newly-determined mass
values and the mass values from Ref. \cite{Kank2006} substituted into the 
AME2003 mass data. The reaction flows shown are more than 10\% 
(thick line) and 1--10\% (thin line) of the reaction flow through the
 triple-$\alpha$-reaction.}
\end{figure}

Figures \ref{fig:flux-trp8} and \ref{fig:flux-jytr} 
show the time-integrated reaction flows relative to the $3\alpha$-reaction
employing the mass data from AME2003 only and the mass data from AME2003
with the addition of the new results, respectively.
In the case of reaction rates employing the newly-determined mass values, the
main reaction flow to nuclei with $A>88$
proceeds through $^{87}$Mo(n,p) $^{87}$Nb(p,$\gamma$) 
$^{88}$Mo(p,$\gamma$) $^{89}$Tc.
In comparison, for reaction rates based on the AME2003 masses
the flow through $^{87}$Mo(p,$\gamma$)$^{88}$Tc is not suppressed.
This opens up two additional paths for significant reaction flows:
$^{88}$Tc(n,p)$^{88}$Mo(p,$\gamma$)$^{89}$Tc
and
$^{88}$Tc(p,$\gamma$)$^{89}$Ru(n,p)$^{89}$Tc.
In consequence, the reaction flow of the previous case,
$^{87}$Mo(n,p)$^{87}$Nb(p,$\gamma$)$^{88}$Mo(p,$\gamma$)$^{89}$Tc,
becomes weaker. When examining the reaction flows 
(Figs. \ref{fig:flux-trp8} and \ref{fig:flux-jytr})
in detail one notices a flow through
$^{87}$Nb(n,p)$^{87}$Zr in the case of the newly-determined mass values 
(see Figure \ref{fig:flux-jytr}).
This is a consequence of a stronger reaction flow through
$^{87}$Mo(n,p)$^{87}$Nb and thus a higher abundance of $^{87}$Nb
compared to the calculation using the AME2003 masses only.
This is seen in the final abundances after decay as increased abundance
of $^{87}$Sr (see Fig.\ \ref{fig:ay}).
Similarly, the relative suppression of the flow through 
$^{88}$Tc(p,$\gamma$)$^{89}$Ru(n,p)$^{89}$Tc
in the calculation using the newly-determined mass values of this work
leads to the observed lower abundance of $^{89}$Y.
The change in the mass of $^{90}$Tc leads to a slight increase in 
the reaction rate for $^{90}$Tc($\gamma$,p)$^{89}$Mo, thus shifting
some mass of the $A=90$ chain into the $A=89$ chain.
As a result, the final abundance of $^{90}$Zr is lower for the calculation
employing the newly-determined mass values of this work.

When comparing the reaction flows in Fig.\ \ref{fig:flux-trp8}
and Fig.\ \ref{fig:flux-jytr} it seems that the newly-determined mass values
allow an additional flow through $^{92}$Rh(p,$\gamma)^{93}$Pd(n,p)$^{93}$Rh.
However, this is an artefact due to the discrete bins of the reaction flows
for plotting purposes.
For the calculation based on the AME2003 mass data these flows are 
just slightly below the bin threshold 
(1\%~of the flow through the $3\alpha$-reaction).
The total flow reaching $^{94}$Pd is very similar in both cases,
only the relative strength in the two paths,
$^{92}$Rh(p,$\gamma$)$^{93}$Pd(n,p)$^{93}$Rh 
and $^{92}$Rh(n,p)$^{93}$Rh(p,$\gamma$)$^{93}$Rh is different.
%
%%%%%%%%%%%%%%%%%%%%%%%%%%%%%%%%%%%%%%%%%%%%%%%%%%%%%%%%%%%%%%%%%%%%%%%%%%%%%%%%%%%%%%%%%%%%%%%%%
%%%%%%%%%%%%%%%%%%%%%%%%%%%%%%%%%%%%%%%%%%%%%%%%%%%%%%%%%%%%%%%%%%%%%%%%%%%%%%%%%%%%%%%%%%%SUMMARY
%%%%%%%%%%%%%%%%%%%%%%%%%%%%%%%%%%%%%%%%%%%%%%%%%%%%%%%%%%%%%%%%%%%%%%%%%%%%%%%%%%%%%%%%%%%%%%%%%%
\section{\label{sum}Conclusions}
The mass values of 21 very neutron-deficient nuclides in the vicinity of the rp- and the 
$\nu$p-process pathways were determined by SHIPTRAP/GSI-Darmstadt and JYFLTRAP/Jyv\"askyl\"a. The 
Penning trap data obtained at both facilities agree within their uncertainties and their weighted 
averages were given as the final results. For more than half of the data these are the first 
experimental determinations. A comparison with the data of the Atomic Mass Evaluation AME2003 and 
a detailed discussion of these and previous experimental results has been conducted for all nuclides. Most previous 
experimental determinations were stemming from $\beta$-endpoint measurements and large modifications 
of the mass surface are observed, in particular when approaching the $N = Z$ line, where experimental 
deviations along isobaric decay chains are accumulated and continued in the extrapolations. Now, the nuclear 
masses necessary for nuclear reaction-rate calculations are based to a 
large extent on experimental data. In addition, the new masses will improve the extrapolations for nuclei further away from stability.\\
The new data were used to perform calculations of the $\nu$p-process nucleosynthesis in order to check their 
impact on the reaction flow and the final abundances. To this end, reaction rates as well as $Q$ values were 
re-calculated and have been compared with results employing the data from AME2003. The detailed flow patterns 
were established and compared. The new mass value for $^{88}$Tc results in a proton-separation energy which is 
$\sim 1$~MeV smaller than in the AME2003 systematics. Consequently, the reaction flow around $^{88}$Tc is 
strongly modified. However, the final abundances for the $\nu$p-process calculations are almost
unchanged.\\
In order to improve the detailed modeling of the reaction pathways even further, a continuation of the 
measurement programme is desired as follows: among the nuclides already studied open questions of ambigous 
mass-to-state assignments should be revisited. These include the nuclides $^{84}\mbox{Y}$, $^{85}\mbox{Nb}$, $^{88}$Tc, 
$^{90}$Tc, and $^{92}$Rh. 
Furthermore, specific nuclides with uncertain spin assignments and/or long-lived isomeric states 
should be extensively searched for to assess their possible influence in the rp- or $\nu$p-process \cite{Novi2001}. 
Eventually new measurements should aim at a study further away from stability towards nuclei at the $N = Z$ line.    
%%%%%%%%%%%%%%%%%%%%%%%%%%%%%%%%%%%%%%%%%%%%%%%%%%%%%%%%%%%%%%%%%%%%%%%%%%%%%%%%%%%%%%%%%%%%%%%%%%
%%%%%%%%%%%%%%%%%%%%%%%%%%%%%%%%%%%%%%%%%%%%%%%%%%%%%%%%%%%%%%%%%%%%%%%%%%%%%%%%%%ACKNOWLEDGEMENTS
%%%%%%%%%%%%%%%%%%%%%%%%%%%%%%%%%%%%%%%%%%%%%%%%%%%%%%%%%%%%%%%%%%%%%%%%%%%%%%%%%%%%%%%%%%%%%%%%%%
\begin{acknowledgments}
We acknowledge financial support by the EU within the networks
NIPNET (contract HPRI-CT-2001-50034), Ion Catcher (contract
HPRI-CT-2001-50022), EURONS (Joint Research Activities TRAPSPEC
and DLEP), the Academy of Finland under the Finnish Centre of
Excellence Programmes 2000-2005 (Project No. 44875, Nuclear and
Condensed Matter Physics Programme) and 2006-2011 (Nuclear and
Accelerator Based Physics Programme at JYFL), the Finnish-Russian Interacademy Agreement (Project \# 8), 
the German Ministry for Education and Research (BMBF contracts: 06GF186I, 06GI185I, 06MZ215), and the Association
of Helmholtz Research Centers (contracts: VH-NG-033, VH-NG-037). This work was supported in part by the Swiss NSF 
(grant 2000-105328). C.F. acknowledges support from an Enrico Fermi Fellowship.
\end{acknowledgments}
\newpage 
%%%%%%%%%%%%%%%%%%%%%%%%%%%%%%%%%%%%%%%%%%%%%%%%%%%%%%%%%%%%%%%%%%%%%%%%%%%%%%%%%%%%%%%%%%%%%%%%%%
%%%%%%%%%%%%%%%%%%%%%%%%%%%%%%%%%%%%%%%%%%%%%%%%%%%%%%%%%%%%%%%%%%%%%%%%%%%%%%%%%%%%%%BIBLIOGRAPHY
%%%%%%%%%%%%%%%%%%%%%%%%%%%%%%%%%%%%%%%%%%%%%%%%%%%%%%%%%%%%%%%%%%%%%%%%%%%%%%%%%%%%%%%%%%%%%%%%%%

\end{document}